\pdfoutput=1
\documentclass[aps,prd,reprint,superscriptaddress,nofootinbib]{revtex4-2}
\usepackage{amsmath}
\usepackage{amsfonts}
\usepackage{amssymb}
\usepackage{graphicx}
\usepackage{bm}
\usepackage{subcaption}
\usepackage{hyperref}
\usepackage{braket}
\usepackage{placeins}

\begin{document}

\title{On chaos in QFT}

\author{Nadav Shrayer}
\email{nadavshraier@tauex.tau.ac.il}
\affiliation{The Raymond and Beverly Sackler School of Physics and Astronomy, Tel Aviv University, Tel Aviv, Israel}

\author{Jacob Sonnenschein}
\email{cobi@tauex.tau.ac.il}
\affiliation{The Raymond and Beverly Sackler School of Physics and Astronomy, Tel Aviv University, Tel Aviv, Israel}

\begin{abstract}
In this note we explore the chaotic behavior of   non-integrable QFTs and compare them to integrable ones. 
We choose as prototypes the  double sine-Gordon   and the sine-Gordon models. We  analyze their discrete spectrum determined by a  truncation method.  We examine the map of the corresponding  energy eigenvalues to  the eigenvalues of  the random matrix theory (RMT)  gaussian orthogonal ensemble (GOE). This is done   by computing the following properties: (a) The distribution of the adjacent  spacings  and their ratios  (b) Higher order spacings and ratios (c) Pair correlations (d) Spectral form factors and (e) Spectral rigidity.  For these properties we  determine the differences between the integrable and non-integrable theories and verify that the former admits a Poisson behavior and the latter GOE (apart from the spectral rigidity, which shows mixed behavior).
\end{abstract}
\maketitle

\preprint{arXiv:2506.10784}

 
\section{Introduction}
Chaotic behavior in classical and quantum mechanical
systems both of a single body and of many bodies has been intensively researched. The study of
ergodicity and chaos in continuous quantum field theories
and string theories is much less mature.

Significant   progress in the study  of quantum chaos in  QFTs in 1+1 dimension has been made  in  \cite{Brandino}, \cite{Prosen} and  \cite{Katz}. In the second a Hamiltonian truncation method was applied to the integrable sine-Gordon (SG) and non-integrable   double sine-Gordon (DSG)   models to determine their  discrete spectrum and in the third a light-cone conformal truncation was used in the context of the 1+1 dimension $\phi^4$ theory. In these papers various aspects of the map between these systems and random matrix theory (RMT) were studied.
 Another  contemporary approach to chaos in QFTs  deals with a quantum analog of the classical Lyapunov exponent, called the Out of   Time-Ordering Correlators (OTOC), and a bound on the maximum value it can hold. Maximally chaotic systems are said to saturate this bound \cite{BHChaos}.
The evolution of systems from integrability to chaos has been intensively studied also using  the method of   “Krylov complexity” (\cite{Rabinovici},\cite{Avdoshkin}\cite{Balasubramanian} and references therein).  

In this note we continue the program of examining the chaotic behavior of QFTs. There are several different directions that one can  follow for that purpose:  (i) Further study of the map between the  discrete spectrum associated with truncated field theories and RMT. (ii) Analyzing chaotic behavior via this map in  different  field theories in different space-time dimensions, like theories of  scalar fields, spinor fields, gauge theories like QED and QCD etc. (iii) Exploring chaotic scattering processes similar to those discovered recently and which   involve high excited string states \cite{Gross:2021gsj},\cite{Rosenhausstrings},\cite{Bianchi:2022mhs}, \cite{Cobistrings},\cite{CobScFF}.
(iv) Computing OTOC and general correlation functions. (v) Using the Krylov complexity method.

Here we follow (i) and present a   followup work to   the paper \cite{Prosen} by
studying properties related to chaos in the SG and DSG models. \footnote{ Contradicting the common lore in \cite{Negro:2022hno}  analytical and numerical evidences was presented  that classical integrable models
possessing infinitely many degrees of freedom, unexpectedly exhibit some features that are typical
of chaotic systems.}

The RMT properties that we study are:
(a) The distribution of the adjacent spacings and their
ratios (b) Higher order spacings and ratios (c) Pair correlations (d) Spectral form factors and (e)
Spectral rigidity.\par

The structure of this paper is as follows: 
In section \ref{Measures} we present several measures of chaos: (i) Adjacent spacings and their ratios (ii)  Higher order spacings and spacing ratios (iii) Pair correlations (iv) Spectral form factor (v) higher order correlations - OTOCs. In section (\ref{DNN})  the distributions of NN spacings and spacings ratios are been determined. 
Section (\ref{Dona}) deals with the distributions of non-adjacent spacings and their ratios. In particular we discuss a probabilistic approach to higher order level spacings. In section (\ref{Pc}) we write down the results of the   pair correlations. The scattering form factor is determined in section (\ref{Sff}).  In section (\ref{SR}) we present the $\Delta_3$ statistics. We conclude with section (\ref{conclusion}) where a summary of the results is given and future directions are discussed. 
We add three appendices to the paper. In appendix (\ref{TCSA}) we review the method of TCSA and the specific parameters used to obtain the data we worked on. In appendix (\ref{SG-and-DSG}) we write down the basic properties of the SG and the DSG models and appendix (\ref{RMT}) is a brief review of random matrix theory.

\section{Measures of chaos} \label{Measures}

One possible way to define a chaotic behavior of a quantum system is if  its energy eigenvalues can be mapped to the eigenvalues of a RMT.  To determine the quality of the map one can compute statistical properties of the energy eigenvalues and compare them to those of RMT. These properties include : (a) The distribution of the adjacent  spacings  and their ratios  (b) Higher order spacings and ratios (c) Pair correlations (d) Spectral form factors and (e) Spectral rigidity. Prior to computing these quantities we elaborate in this section on each of them. We begin with discussing the unfolding of the spectrum, which is a necessary process in order to compare different spectra.

\subsection{Unfolding}
The universality of chaos is often seen through local statistics of the spectrum, while longer ranged statistics are affected by the model dependent averaged level density $\rho_0$, so it is required to decouple the spectrum from its averaged level density in order to make meaningful comparisons between spectra. This is done in a process called "unfolding", where the average level density is 'flattened' such that it is constant throughout the spectrum, and only the fluctuation around it remain. Another model dependent aspect of the spectra is the scale of the energy levels, so the unfolding should also align all spectra to the same local scale, which is naturally chosen to be the mean nearest neighbor spacing $\langle s\rangle$ such that after unfolding it is equal to 1.\par
A common scheme for level unfolding is the use of the cumulative distribution of the normalized average level density, where the unfolded levels are $N$ times the value of the CDF

\begin{equation}
\label{eq:Unfolding}
e_i = N \int_{-\infty}^{E_i} \rho_{0} (E') dE'
\end{equation}

To see the effect of unfolding explicitly, we'll use an example from RMT. Assume $\{ \lambda _i \}$ is a set of $N$ eigenvalues taken from a GUE and normalized by $1/\sqrt{N}$ . The GUE average level density $\rho_0$ is found by using $\beta=2$ in (\ref{eq:rho_0})

\begin{equation}
\label{eq:rho_0-GUE}
\rho_0(\lambda) = \frac{1}{2\pi} \sqrt{4-\lambda^2}
\end{equation}

\begin{figure}[tbp]
     \centering
   \begin{subfigure}{\columnwidth}
         \centering
        \includegraphics[width=\linewidth]{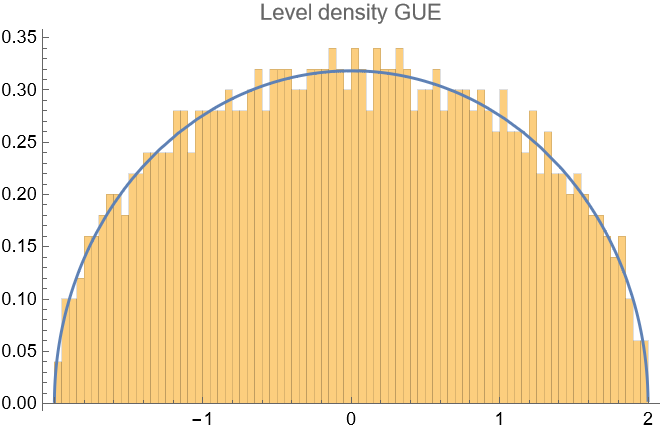}
         \caption{}
         \label{fig:GUE-N=1000-hist}
     \end{subfigure}
   \begin{subfigure}{\columnwidth}
         \centering
        \includegraphics[width=\linewidth]{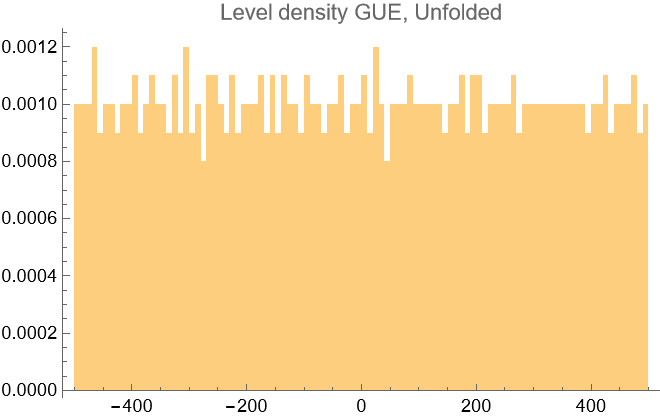}
         \caption{}
         \label{fig:GUE-N=1000-hist-unfolded}
     \end{subfigure}
            \caption{Histogram of the spectrum taken from GUE, $N=1000$. The blue curve in (\ref{fig:GUE-N=1000-hist}) is the average level density of GUE, eq. (\ref{eq:rho_0-GUE}) and it is used for the unfolding. (\ref{fig:GUE-N=1000-hist-unfolded}) shows the spectrum after the unfolding.}
        \label{fig:total-GUE-N=1000}
\end{figure}

If we compute the set of eigenvalues numerically with $N=1000$, we can see in fig (\ref{fig:GUE-N=1000-hist}) that the histogram of eigenvalues (the orange bars) matches the average level density in (\ref{eq:rho_0-GUE}), and the mean value of the nearest neighbor spacings calculated numerically from GUE, and in fig (\ref{fig:GUE-N=1000-hist-unfolded}) we can see the histogram of the set of levels $\{ e_i \} $ after applying the unfolding procedure in (\ref{eq:Unfolding}) using (\ref{eq:rho_0-GUE}). As can be seen, the unfolded spectra 'flattens' $\rho_0$ to a constant, while preserving the fluctuations around it. The mean nearest neighbor spacing turns out to be $1.00026$. Note that since $\rho_0$ is a normalized distribution (\ref{eq:Unfolding}) the range of the spectrum changes such that $e_{max} - e_{min}=N$.
 
\begin{figure} [tbp]
     \centering
   \begin{subfigure}{\columnwidth}
         \centering
        \includegraphics[width=\linewidth]{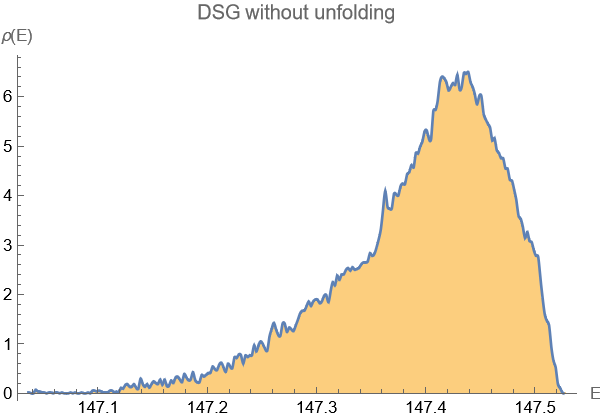}
         \caption{DSG spectrum before unfolding}
         \label{fig:DSG-spectrum-before-unfolding}
     \end{subfigure}
   \begin{subfigure}{\columnwidth}
         \centering
        \includegraphics[width=\linewidth]{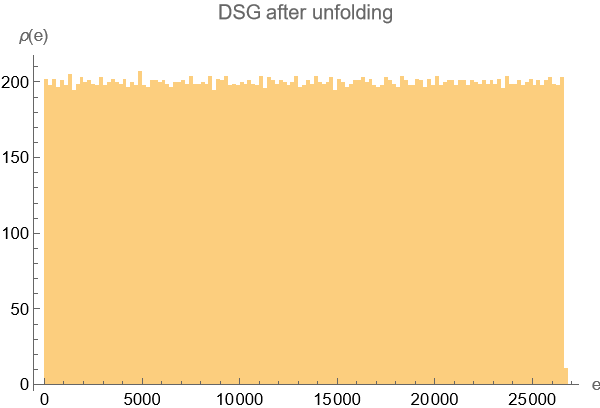}
         \caption{DSG spectrum after unfolding}
         \label{fig:DSG-spectrum-after-unfolding}
     \end{subfigure}
            \caption{The high range spectrum of DSG with $mL=1$. The orange bars in (\ref{fig:DSG-spectrum-before-unfolding}) is the binned density and the blue curve on top is the interpolated $\rho_0$. (\ref{fig:DSG-spectrum-before-unfolding}) shows the same spectrum after unfolding, using the numerically evaluated $\rho_0$. The fluctuations are still visible in (\ref{fig:DSG-spectrum-before-unfolding}) as variations of the bins heights around the constant average density.}
        \label{fig:DSG-unfolding}
\end{figure}

When the average level density is not known, it can be calculated numerically using the histogram of the spectrum, where the bin sizes should be larger than the mean spacing but much smaller than the scale where $\rho_0$ changes. For example, we take the high-range of the DSG spectrum with $mL=1$ and get a histogram with properly sized bins. We use the bin counts and average energy of each bin as data points and interpolate a smooth function in between, which is our numerical $\rho_0 (E)$, which we then use for the unfolding (see fig. \ref{fig:DSG-unfolding}).

Notice that the energy scale of the unfolded spectrum changed, and the whole spectrum is now between 0 and $N$.
In areas where the derivative of $\rho_0$ is large, more points should be used in the interpolation, although one should bear in mind that using too many points might flatten the fluctuations as well. One should also be careful not to get negative values for the interpolated $\rho_0$ since those could cluster the energy levels within the energy range where $\rho_0$ is negative, once unfolded.

\subsection{Adjacent  spacings and their  ratios}

An important  step in the analysis of  quantum chaos was made in  the work of Wigner during the 1960's, when he proposed that the statistics of the eigenvalues and eigenstates of heavy nuclei and complex molecules could be understood in terms of Random Matrix Theory, where the Hamiltonian of the system is replaced by an IID and rotationally invariant random matrix, and the energy levels of the system are given by the eigenvalues of the matrix.
In this way it is possible to compute, using random matrices, different observables seen in experiments. Wigner and Dyson\cite{Dyson:1962b} observed that the spacings between consecutive energy levels obey a kind of distribution that could be derived analytically from RMT by diagonalizing $2X2$ Gaussian matrices. The distribution obtained is called the Wigner surmise, or Wigner-Dyson (WD) distribution, and it is

\begin{equation}
\label{eq:WD-distribution}
P_{\beta}(s) = A_{\beta} s^{\beta} e^{-B_{\beta}s^2}
\end{equation}

Where $s$ is the spacing between two consecutive energy levels (or eigenvalues in terms of RMT): $$s_n=E_{n+1}-E_{n}$$ $\beta$ is the Dyson index\cite{Dyson:1962b} and $A_{\beta}, B_{\beta}$ are constants explicitly shown in table (\ref{tab:gaussian-ensembles}). They are determined by the requirement that the distribution in (\ref{eq:WD-distribution}) must be normalized and it's mean must be equal to 1.

\begin{table}[tbp]
\centering
\begin{tabular}{c c c}
\hline\hline
Ensemble & $\beta$ & $(A_\beta,\;B_\beta)$ \\
\hline
GOE & $1$ & $\left(\frac{\pi}{2},\,\frac{\pi}{4}\right)$ \\
GUE & $2$ & $\left(\frac{32}{\pi^2},\,\frac{4}{\pi}\right)$ \\
GSE & $4$ & $\left(\frac{262144}{729\pi^3},\,\frac{64}{9\pi}\right)$ \\
\hline\hline
\end{tabular}
\caption{Constants for the three Gaussian ensembles.}
\label{tab:gaussian-ensembles}
\end{table}

We note that the WD distribution is an approximation of the consecutive level spacing probability, which should be derived from a general $N\times N$ random matrices.
Bohigas, Giannoni and Schmidt conjectured \cite{BGS} in the 80's that the local statistics of energy levels of chaotic systems are universal, and depend only on the symmetries of the system, which correspond to Gaussian (or circular, for this matter) ensembles in RMT with the same local statistics: systems with time-reversal symmetry to the Gaussian orthogonal ensemble (GOE) , systems without time reversal symmetries to the Gaussian unitary ensemble (GUE), and systems with time reversal symmetry but spin $1/2$ statistics with to Gaussian symplectic ensemble (GSE). They also posited that for integrable systems, the Poisson ensemble captures the local statistics. This follows Berry and Tabor conjecture \cite{BerryTabor} that all integrable systems have uncorrelated energy levels. From the Poisson ensemble, which is an IID diagonal random matrix where the eigenvalues are random variables. The consecutive level spacing distribution derived from the Poisson ensemble is

\begin{equation}
\label{eq:Poisson-distribution}
P_{PE}(s) = e^{-s}
\end{equation}

For systems in a mixed state of integrability and non-integrability, there are several approaches to describing the distributions of consecutive level spacings. We will describe two approaches, both phenomenological. The first example is due to Brody \cite{Brody1,Brody2} which uses a heuristic argument by Wigner and an ansatz for a repulsion factor that interpolates between the PE and GOE to obtain the Brody distribution

\begin{equation}
\label{eq:Brody-distribution}
P_{q}(s) = (q+1)a_{q}s^{q}\exp{\left(-a_{q}s^{q+1}\right)}
\end{equation}

Where $q$ is an interpolating parameter that lies in the range $0\leq q \leq 1$ and $$a_{q}=\left[ \Gamma \left(\frac{q+2}{q+1}\right)\right]^{q+1}$$ is obtained by normalizing the distribution and requiring a mean of 1. $\Gamma$ is the gamma function. Note that when $q=0$ the distribution is exactly (\ref{eq:Poisson-distribution}), and for $q=1$ it is the GOE nearest neighbor spacing (\ref{eq:WD-distribution}) with $\beta=1$. The Brody distribution does not interpolate to the other Gaussian ensembles.\par
Another interpolation, this time based on localization and due to Izraelev \cite{Izrailev1,Izrailev2} is

\begin{equation}
\label{eq:Izraelev-distribution}
P_{\nu}(s) =A_{\nu}s^{\nu}\exp{\left(-\frac{\pi^2}{16}\nu s^2-\left(B_{\nu}-\frac{\nu}{2}\right)\frac{\pi}{2}s\right)}
\end{equation}

Where, again, $A_{\nu},B_{\nu}$ are obtained by normalizing the distribution and its mean to 1. The parameter $\nu$ now holds for the whole range $0\leq \nu \leq 4$, and for the specific values $\nu=0,1,2$ it deviates from (\ref{eq:Poisson-distribution}), and (\ref{eq:WD-distribution}) for GOE, GUE respectively by at most $5\%$. We note that both $\nu$ and $q$ are phenomenological parameters, and do not have a direct physical meaning.\par
In recent years there is a use of another statistical measure of local correlations in the energy levels, due to Oganesyan and Huse \cite{OganesyanHuse} - the consecutive spacing ratios $r_n$, defined as

\begin{equation}
\label{eq:NN-spacing-ratio}
r_{n}=\frac{s_{n+1}}{s_n}=\frac{E_{n+2}-E_{n+1}}{E_{n+1}-E_{n}}
\end{equation}

Atas, Bogomolny, Giraud and Roux \cite{ABGR} used a derivation similar to that which gave rise to the WD distribution only this time using $3\times3$ random matrices since (\ref{eq:NN-spacing-ratio}) requires at least 3 eigenvalues, and got the following approximate formula for the distribution of nearest neighbor (NN) level spacing ratios

\begin{equation}
\label{eq:NN-spacing-ratio-distribution}
P_{\beta}(r)=\frac{1}{N_\beta}\frac{(1+r)^{\beta}}{(1+r+r^2)^{1+3\beta/2}}
\end{equation}

Where $N_\beta$ is a normalization constant assuring the distribution above is normalized to unity.
As for the WD distribution, a better approximation for the distribution would use bigger matrices. Atas, Bogomolny, Vivo and Vivo derive analytical expression of such distributions for up to $5\times5$ matrices \cite{ABGVV}, and a general prescription for an $N \times N$ expression, but comparison with numerical data shows that (\ref{eq:NN-spacing-ratio-distribution}) is already a very good approximation.\par
One of the main reasons for the use of the spacing ratios rather than the spacings is the assumption that nearest neighbor levels are still in the scale in which the mean level density $\rho_0$ is constant and can be factored out, and so in taking the spacing ratio they cancel. This means that the level spacing ratios are independent of the mean level spacing and have no need of being unfolded.
Since the order of the spacings in the definition of the spacing ratio (\ref{eq:NN-spacing-ratio}) was chosen at random and could have also been $r_{n}=s_{n}/s_{n+1}$ which is the inverse of the original definition, the whole support of the spacing ratios is sometimes restricted to being from 0 to 1 by the defining

\begin{equation}
\label{eq:NN-spacing-ratio-tilde}
\tilde{r}_{n}=\min \Bigl\{ r_{n},\frac{1}{r_n} \Bigr\}
\end{equation}

A consecutive level spacing ratio distribution was also calculated in \cite{ABGR} for the integrable case based on (\ref{eq:Poisson-distribution}) for $3\times3$ matrices by assuming that the joint probability distribution is the product of independent probabilities. This leads to the expression

\begin{equation}
\label{eq:NN-spacing-ratio-distribution-Poisson}
P_{PE}(r)=\frac{1}{(1+r)^2}
\end{equation}

\subsection{Higher order spacings and spacing ratios} \label{higher-orders}

It is a natural generalization to look at the next orders in the level spacings and spacing ratios in order to better understand the transition between short range and long range correlations in the level statistics, the latter will be dealt with in the next section.\par
In this section we will denote $k$-spacing as

\begin{equation}
\label{eq:k-spacing}
s_{n,k} = E_{n+k}-E_n
\end{equation}

Where the nearest neighbor spacing is the case when $k=1$. In the literature the $k$-spacings are sometimes noted such that nearest neighbor spacings have $k=0$, so careful read is needed.
There are several approaches for analytical derivations of the $k$-spacing probability $P_{k}(s)$ \cite{ChenuMartinezShir}. Fox, Kahn \cite{FoxKhan} derive a series expansion in the spacing $s$ directly from RMT. Other more direct approaches involve a calculation similar to that carried by Brody \cite{Brody1} which is based on a statistical argument originally made by Wigner, where the spacing probability $P(s)$ is derived from the form of the level repulsion at small $s$ via a heuristic repulsion function $r(s)$ \cite{EngelMainWunner}. 

\begin{equation}
\label{eq:repulsion-function-spacing}
P(s) = r(s)\exp{\left(-\int_0^s r(s')ds'\right)}
\end{equation}

Abu Magd and Simbel \cite{AbulMagdSimbel} use the ansatz $r(k,s)\propto s^{k\beta}$ and a recursion relation to solve for the level repulsion of of the $k$-spacing. Adding a Gaussian tail for the proper large s behavior, they obtain a WD like expression

\begin{equation}
\label{eq:Abu-magd-k-spacing-probability}
P_{k}(s) = A_{k,\beta}s^{\alpha_{k,\beta}}\exp{\left(-B_{k,\beta}s^2\right)}
\end{equation}

Where

\begin{equation}
\label{eq:Abu-magd-k-spacing-probability-exponent}
\alpha_{k,\beta}=\frac{k(k+1)}{2}\beta+(k-1)
\end{equation}

and $A_{k,\beta}$ and $B_{k,\beta}$ are determined by proper normalization. The same expression (\ref{eq:Abu-magd-k-spacing-probability}) is obtained in \cite{Rao} by inserting the $k$-spacing as a delta function to the Gaussian joint probability density, and evaluating it directly. \par
As for $k$-spacing ratios, it is conjectured in \cite{HarishniTekur} that the $k$-spacing ratio probability distribution is the same as (\ref{eq:NN-spacing-ratio-distribution}), but replacing $\beta$ by the expression $\alpha_{k,\beta}$ given in (\ref{eq:Abu-magd-k-spacing-probability-exponent}).\par
There is also an explicit analytical derivation of $k=2$ spacing ratio distribution given in \cite{ABGVV}.

\subsection{Pair correlations} \label{pair-correlations}

The NN spacing distribution and the higher order $k$-spacing generalization are related to another concept - the pair correlation function and the pair cluster function, which are the $n=2$ cases of the general $n$-point correlation and cluster functions, discussed in appendix \ref{RMT}. We begin with the $n$-point correlation function (\ref{eq:n-point-correlation}), which is the the $n$ marginal probability distribution of the joint probability distribution of the ensemble, and choose to work in the GUE case in the following part for simplicity, but a full derivation for all three symmetry ensemble can be found in \cite{Weidenmuller}. Using determinant identities, we can insert the Gaussian measure into the Vandermonde determinant by a polynomial reparameterization of the entries of using the normalized harmonic oscillator eigenvectors:

\begin{equation}
\label{eq:normalized-Hermite-polynomials}
\phi_n(x)=\frac{1}{\sqrt{\sqrt{\pi}2^{n}n!}}e^{-\frac{1}{2}x^2}H_n (x)
\end{equation}

Where $H_n (x)$ are the regular Hermite polynomials

\begin{equation}
\label{eq:regular-Hermite-polynomials}
H_n(x)=(-1)^n e^{x^2}\frac{d^n}{dx^n}e^{-x^2}
\end{equation}

We can define the kernel

\begin{equation}
\label{eq:Hermite-kernel}
K_N(\lambda_1,\lambda_2)=\sum_{n=0}^{N-1} \phi_n(\lambda_1)\phi_n(\lambda_2)
\end{equation}

and check, using the orthogonality of the Hermite polynomials, that it satisfies the kernel equation

\begin{equation}
\label{eq:kernel-equation}
\int K_N(\lambda_1,\lambda_2)K_N(\lambda_2,\lambda_3)d\lambda_2=K_N(\lambda_1,\lambda_3)
\end{equation}

With the kernel we can write the $n$-point correlation function as

\begin{equation}
\label{eq:n-correlation-kernel}
R(\lambda_1,\lambda_2,\dots,\lambda_n)=\det\left[K_N(\lambda_i,\lambda_j)\right]_{i,j=1,2,\dots,n}
\end{equation}

Now unfolding the spectrum allows us to simplify this expression further, since unfolding turns the mean level spacing, which is the 1-point correlation function, into a constant $R(\lambda)=\rho_{0}(\lambda)=1/D$ where $D$ is the mean NN level spacing, which is usually taken to be unity in the unfolding process. From this follows that the correlation between any to levels is translational invariant, i.e. depends only on expressions like $\lambda_i-\lambda_j$. Using rescaled eigenvalues $x=D\lambda$ and the large $N$ limit, we get the sine kernel

\begin{equation}
\label{eq:sine-kernel}
\lim_{N\to\infty} DK_N(x_i,x_j) = \frac{\sin(\pi(x_i-x_j))}{\pi(x_i-x_j)}
\end{equation}

And for the 2-point GUE correlation function we get the expression

\begin{equation}
\label{eq:2-point-correlation-GUE}
R(x_1,x_2)=1-\left(\frac{\sin(\pi s)}{\pi s}\right)^2
\end{equation}

Where the 1 comes from the the disconnected part $R(x_1)R(x_2)$ which where normalized by choosing $D=1$, and the second term is the connected part, which is the 2-point cluster function $T(x_1,x_2)$ with $s=x_2-x_1$.
The 2-point correlation function for GOE is derived in the same manner, though (\ref{eq:n-correlation-kernel}) is more complicated. We will write it when it is used, in section \ref{Pc}.\par
The pair correlation after unfolding can be seen as the probability distribution for having a spacing $s$ between any two eigenvalues in the spectrum, and so we can relate it to the higher order spacing distributions

\begin{equation}
\label{eq:2-point-correlation-and-k-spacings}
R(x_1,x_2)=\sum_{k=1}^{N-1}P_{k}(s)
\end{equation}

One known use of the pair-correlation is found in the surprising connection between the Riemann hypothesis (RH) and the the GUE. As a reminder, the RH conjectures that the non-trivial zeros $z_n$ of the Riemann zeta function $\zeta(s)=\sum_n^\infty n^{-s}$ are all found along the line $\Re\{s\}=1/2$

\begin{equation}
\label{eq:Riemann-zeros}
z_n=\frac{1}{2}+i\theta_n
\end{equation}

Montgomery conjectured \cite{Montgomery}, based on the RH, an estimate for the correlation between any two imaginary parts of zeros of the zeta function $\theta, \theta'$. For fixed $0<\alpha<\beta<\infty$

\begin{widetext}
\begin{equation}
\label{eq:montgomery-pair-correlation}
\frac{|\{(\theta,\theta'): 0<\theta,\theta'\leq T, 2\pi\alpha(\log T)^{-1}\leq \theta-\theta'\leq 2\pi\beta(\log T)^{-1} \}]|}{(T \log T)/2\pi}
\sim \int_\alpha^\beta  1-\left(\frac{\sin(\pi u)}{\pi u}\right)^2 du
\end{equation}
\end{widetext}

Where it is assumed that $T\rightarrow\infty$. This was later confirmed numerically in by Odlyzko in \cite{Odlyzko}.\par Dyson \cite{Dyson:1962b} pointed out that this behavior of pair correlations is identical to that found in GUE. Since the pairs in GUE are ordered eigenvalues of a random matrix, this similarity pushes forward the Hilbert and Polya conjectures (see \cite{Montgomery}) that the nontrivial zeros of the zeta function correspond to eigenvalues of a positive linear operator.

\subsection{Spectral Rigidity}
The pair correlation is also present in two closely related statistical measures of spectra - the level number variance $\Sigma^2(L)$ and the spectral rigidity $\Delta_3(L)$, both indicators of short and long range phenomena between pairs of levels. We shall derive both based on \cite{stockmann}.\par
Given an unfolded spectrum, the number of energy levels in a given energy interval of length $L$ is expressed with the level density $\rho(x)$

\begin{equation}
\label{eq:number-of-levels}
n(x,L)=\int_x^{x+L} \rho(x')dx'
\end{equation}

and the mean number of levels $\langle n(L) \rangle$ is the ensemble average of (\ref{eq:number-of-levels}), which is independent of a specific choice of lower bound energy level $x$ because it averages over all of them. Since the unfolding process assures that the mean spacing between NN levels is 1, it follows that the mean number of levels in an interval $L$ is $L$, i.e. $\langle n(L) \rangle=L$.
We can now define the variance in the mean number of levels as

\begin{equation}
\label{eq:mean-number-of-levels-variance}
\Sigma^2(L)=\langle {n(L)}^2 \rangle - {\langle n(L) \rangle}^2
\end{equation}

and need only to calculate the first term, which is explicitly

\begin{equation}
\label{eq:mean-number-of-levels-squared}
\langle {n(L)}^2 \rangle =\iint \langle \rho(x_1)\rho(x_2)\rangle dx_1dx_2
\end{equation}

the term $\langle \rho(x_1)\rho(x_2)\rangle$ is the pair correlation function for $x_1\neq x_2$. For $x_1=x_2$ we get and extra delta function term. Overall $\langle \rho(x_1)\rho(x_2)\rangle=1-T(x_1,x_2)+\delta(x_1-x_2)$, and inserting this into (\ref{eq:mean-number-of-levels-variance}) gives

\begin{equation}
\label{eq:mean-number-of-levels-variance-explicit}
\Sigma^2(L)=L-2\int_0^L (L-s)T(s)ds
\end{equation}

Where $s=x_1-x_2$. Using the pair cluster functions for the GOE and GUE respectively yields the final expressions for ${\Sigma_\beta}^2(L)$, which asymptotically in the $L\rightarrow\infty$ limit are

\begin{equation}
\label{eq:mean-number-of-levels-variance-asymptotics}
\begin{split}
    {\Sigma_1}^2(L)&= \frac{2}{\pi^2} \left( \ln(2\pi L) + \gamma+1-\frac{\pi^2}{8}\right)  \\
    {\Sigma_2}^2(L)&= \frac{2}{\pi^2} \left( \ln(2\pi L) + \gamma+1\right) 
\end{split}
\end{equation}

Where $\gamma$ is the Euler-Mascheroni constant and we use $\beta$ as the Dyson index in the subscript.\par
The spectral rigidity $\Delta_3(L)$ is the minimal variance of the difference between the number of levels in an interval $L$, $n(x,L)$, and straight line, averaged over the spectrum.

\begin{equation}
\label{eq:spectral-rigidity}
\Delta_3(L)=\frac{1}{L} \Biggl \langle  \min_{A,B} \int_x^{x+L} \left( n(x,x')-Ax'-B) \right)^2dx' \Biggr \rangle
\end{equation}

which can be expressed explicitly as

\begin{equation}
\label{eq:spectral-rigidity-explicit}
\Delta_3(L)=\frac{L}{15}-\frac{1}{15L^4}  \int_0^{L} \left(L-s\right)^3 \left(2L^2-9Ls-3s^3\right)T(s)ds
\end{equation}

For integrable cases, the 2-point cluster function vanishes and the spectral rigidity is $\Delta_{3,0}(L)=\frac{L}{15}$. For the other ensembles, inserting the right 2-point cluster functions in (\ref{eq:spectral-rigidity-explicit}) and taking the large $L$ limit gives

\begin{equation}
\label{eq:spectral-rigidity-asymptotics}
\begin{split}
   \Delta_{3,1}(L)&= \frac{1}{\pi^2} \left( \ln(2\pi L) + \gamma-\frac{5}{4}-\frac{\pi^2}{8}\right)  \\
    \Delta_{3,2}(L)&= \frac{1}{2\pi^2} \left( \ln(2\pi L) + \gamma-\frac{5}{4}\right) 
\end{split}
\end{equation}

which is very similar to the ${\Sigma_{\beta}(L)}^2$ expressions in (\ref{eq:mean-number-of-levels-variance-asymptotics}).\par 

\subsection{Spectral form factor}
To conclude this part, we note another popular measure related to the pair correlation - the spectral form factor (SFF). It arises naturally when dealing with the time dependence of the expectation value of two local operators in a finite temperature $\beta$ (not to be confused with $\beta$ as the Dyson index), and has been of recent interest in showing the discreteness of the spectrum of black holes using holography from the bulk perspective \cite{BHChaos} and in studying the thermalization of high energy states in the eigenstate thermalization hypothesis (ETH), see \cite{Kafri} for a review on the subject. 
Consider the 2-point correlation function of a Hermitian operator $O(t)$ 

\begin{equation}
\begin{split}
G(t)\
&= \frac{1}{Z(\beta)} Tr\left( e^{-\beta H} O(t)O(0) \right)\\
&= \frac{1}{Z(\beta)} \sum_{m,n} e^{-\beta E_m} {|\langle m|O|n\rangle|}^2 e^{-i(E_m-E_n)t}
\end{split}
\label{eq:pair-correlation-hermitian-operator}
\end{equation}

Where $Z(\beta)=Tr\left(e^{-\beta H}\right)$ is the partition function and $|m\rangle$ is the energy eigenstate with energy $E_m$. The time dependence is mostly characterized by the energy levels, which can be expressed better using the analytically continued partition function $Z(\beta,t)=Tr\left(e^{-(\beta+it)H}\right)$ and writing

\begin{equation}
\label{eq:pair-correlation-SFF}
g(t)=\bigg| \frac{Z(\beta,t)}{Z(\beta)}\bigg|^2 =\frac{1}{{Z(\beta)}^2} \sum_{m,n} e^{-(E_m+E_n)\beta}e^{-i(E_m-E_n)t}
\end{equation}

Which reduces, for infinite temperature ($\beta=0$) and a finite spectrum of size $N$ to the expression

\begin{equation}
\label{eq:discrete-SFF}
g(t)=\frac{1}{{N}^2} \sum_{m,n} e^{-i(E_m-E_n)t}
\end{equation}

In the continuum limit, taking an ensemble mean (\ref{eq:discrete-SFF}) gives

\begin{equation}
\label{eq:continuous-SFF}
\langle g(t) \rangle=\frac{1}{{N}^2} \iint \langle \rho(E_1)\rho(E_2) \rangle  e^{-i(E_1-E_2)t} dE_1dE_2
\end{equation}

where in general we can separate the level density into $\rho(E)=\rho_0(E)+\delta\rho(E)$ such that taking the ensemble average is the mean level density $\langle \rho(E) \rangle=\rho_0(E)$. This gives $$\rho_0(E_1)\rho_0(E_2) + \langle\delta\rho(E_1)\delta\rho(E_2)\rangle$$ which can be inserted into (\ref{eq:continuous-SFF}), resulting in two terms - the first one is the square of the Fourier transform of $\rho_0(E)$, which is a model dependent term that decays for $t\rightarrow\infty$ if the spectrum is finite. This term is usually referred to as the 'descent'. After unfolding, the mean level density is unity, and the second term is the normalized cluster function $\langle\delta\rho(E_1)\delta\rho(E_2)\rangle=\delta(x_1-x_2)-T(x_1,x_2)$. Inserting the proper $T(x_1,x_2)$ terms for the GOE and GUE ensembles gives, after Fourier transformation, the connected part of the SFF \cite{Haake}

\begin{equation}
\label{eq:SFF-connected-GOE}
\langle g_c(t) \rangle_{GOE}=
  \begin{cases}
\delta(t)+2|t|-|t|\log(1+2|t|) & \text{for } |t| \leq 1 \\
2-|t|\log\left(\frac{2|t|+1}{2|t|-1}\right) & \text{for } |t| >1
    \end{cases}
\end{equation}

\begin{equation}
\label{eq:SFF-connected-GUE}
\langle g_c(t) \rangle_{GUE}=
  \begin{cases}
\delta(t)+|t| & \text{for } |t| \leq 1 \\
1 & \text{for } |t| >1
    \end{cases}
\end{equation}

\begin{figure} [tbp]
    \centering
    \includegraphics[width=\columnwidth]{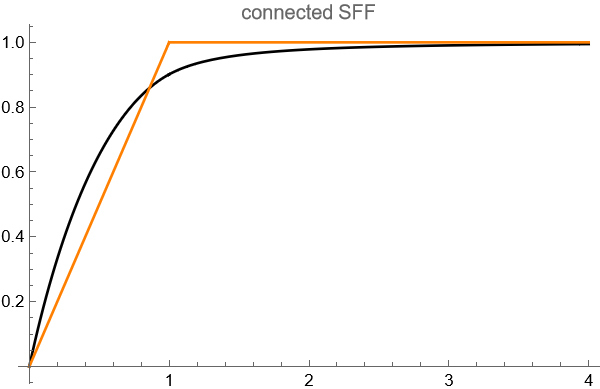}
    \caption{RMT expressions for the connected parts of the SFF, after unfolding and normalization. The black curve corresponds to the GOE (\ref{eq:SFF-connected-GOE}) and the orange curve to the GUE (\ref{eq:SFF-connected-GUE}).}
    \label{fig:connected-SFFs-RMT}
\end{figure}

These terms in the SFF are usually referred to as the 'ramp' for $t<1$ and 'plateau' for $t>1$, as seen for the GOE and GUE SFF in Fig. \ref{fig:connected-SFFs-RMT}. The plateau part can be thought of as the late time average of the SFF $\lim_{T\rightarrow\infty}\frac{1}{T}\int_0^T g(t)dt$ and it is due to the discreteness of the spectrum.
The ramp is associated with the level repulsion, and indeed for integrable systems where the pair cluster function $T(x_1,x_2)$ vanishes there is no ramp in the SFF, and the descent part goes directly to the plateau. It was suggested in the past years that the indication of a ramp in the SFF is a measure for quantum chaos, though there are examples for spectra that give rise to ramps in the SFF even though they could be written as analytical expressions and are therefore not chaotic \cite{SFFramp}.

\section{Distributions of NN spacings and spacings ratios}\label{DNN}

In this section we apply the characteristics of quantum chaos mentioned in the previous section to the a numerically calculated spectrum of the non-integrable DSG model and compare the results with those of the spectrum of SG.\par 
In particular, we would like to affirm that the level statistics behave like those predicted by RMT with the corresponding ensemble being the GOE (or COE, since their local level statistics are the same and they differ only in the average level density), Since DSG is symmetric under time reversal transformation. This goal can be translated into finding that the value of $\beta$ is indeed  $\beta=1$  in the RMT $\beta$ ensemble. If some measurements of the value of $\beta$ deviates from 1, taking into account statistical errors, this might mean that the system is in a transition between two analogous ensembles of RMT. Specifically, if $\beta$ is between 0 and 1 the system might be in a transitional phase between the Poisson ensemble, which is analogous to integrable systems, and GOE, which is non-integrable with time reversal symmetry. Such mix states can be thought of as the quantum version of the classical KAM theory in which integrable and chaotic regions coexist in the phase space.\par
The data used was calculated by Prosen, Srdinsek and Sotiriadis in their paper \cite{Prosen}, they used an energy cutoff via a Hamiltonian truncation method called 'Truncated Hamiltonian Space Approach' (TCSA) in order to calculate the spectrum. The coupling constant $\lambda$ is chosen such that the dimensionless parameter $l=m_{\beta}L$ will be of order 1, where $m_{\beta}$ is the mass scale corresponding to the first breather's mass for a given frequency parameter $\beta$ in the SG potential, and $L$ is the size of the system (i.e. the model has the spatial identification $x \sim x + L$). The two datasets we  use are SG and DSG, both with $l=1$ and containing around $85,000$ energy levels. The SG data with the frequency $\beta=1$ and the DSG with $(\beta_1,\beta_2)=(1,2.5)$, a combination of frequencies that the authors found to introduce the most GOE like behavior, at least in terms of the mean consecutive spacing ratio, $\langle \tilde{r} \rangle$. In the following calculations, we use only the highest (around $25,000$) energy levels out of the whole dataset, which we unfold by interpolation. \par
\begin{figure} [tbp]
     \centering
   \begin{subfigure}{\columnwidth}
         \centering
        \includegraphics[width=\linewidth]{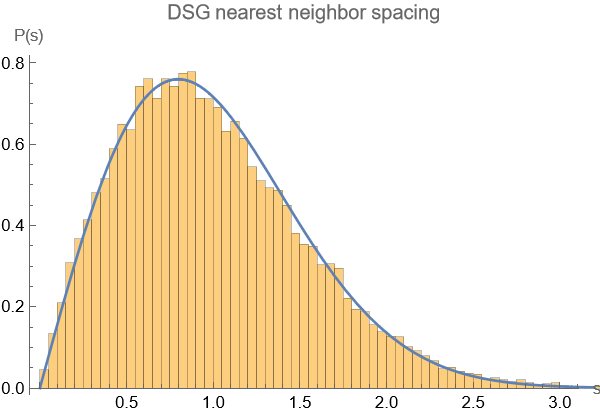}
         \caption{}
         \label{fig:DSG-NN-spacing}
     \end{subfigure}
   \begin{subfigure}{\columnwidth}
         \centering
        \includegraphics[width=\linewidth]{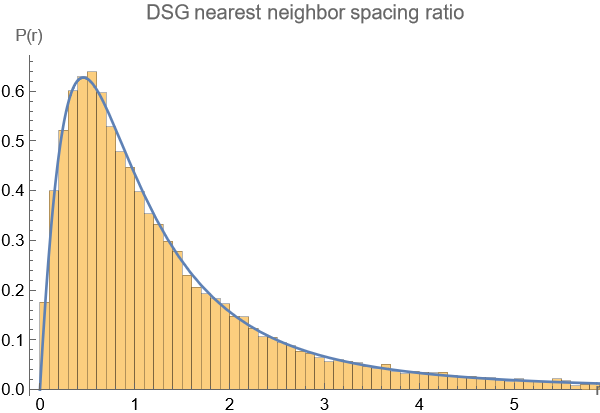}
         \caption{}
         \label{fig:DSG-NN-spacing-ratio}
     \end{subfigure}
            \caption{(\ref{fig:DSG-NN-spacing}) The unfolded DSG nearest neighbor spacing (orange bars) with the GOE Wigner-surmise (blue curve. (\ref{fig:DSG-NN-spacing-ratio}) The unfolded DSG nearest neighbor spacing ratio (orange bars) with the expression for nearest neighbor spacing ratio distribution (\ref{eq:NN-spacing-ratio-distribution}), with $\beta=1$ (blue curve).}
        \label{fig:DSG-NN-spacing-and-spacing-ratio}
\end{figure}

To begin with, We compute the nearest-neighbor spacing distribution, compared with the analtytical Wigner-Dyson distribution for $\beta=1$. The results seem to fit nicely (see fig (\ref{fig:DSG-NN-spacing})). We also compare the nearest nieghbor spacing ratio distribution with the analytical expression, see fig. (\ref{fig:DSG-NN-spacing-ratio}), a calculation not done in the original paper.

For a numerical comparison, we compute the mean value of the nearest neighbor spacing ratio $\langle \tilde{r} \rangle=0.529808$ which is compared with the mean of the RMT based spacing ratio "surmise", given in (\ref{eq:NN-spacing-ratio-distribution}) $\langle \tilde{r} \rangle_{GOE} =0.53590$, which gives a deviation of less than $1\%$ between the two values. We note that this calculation is the basis for the choice of the frequency parameters in DSG, in such a way that the combination of the frequencies $(\beta_1,\beta_2)$ has the value of $\langle \tilde{r} \rangle$ closest to the RMT prediction, but there always seems to be a minimal deviation of the same order from the prediction, indicating that the DSG data is not completely chaotic.\par

Next we would like  to compare  the results above with the corresponding integrable SG model. We use data calculated numerically by the same authors, with the conditions used in DSG that $m_\beta L=1$, and a frequency $\beta=2.5$. As for the DSG spectrum, we take only the highest $\sim 25,000$ energy level as the dataset and unfold it. From the unfolded data we compute the consecutive level spacing and spacing ratio and plot it with using (\ref{eq:Poisson-distribution}) and (\ref{eq:NN-spacing-ratio-distribution-Poisson}) as RMT predictions 

\begin{figure} [tbp]
     \centering
   \begin{subfigure}{\columnwidth} 
         \centering
        \includegraphics[width=\linewidth]{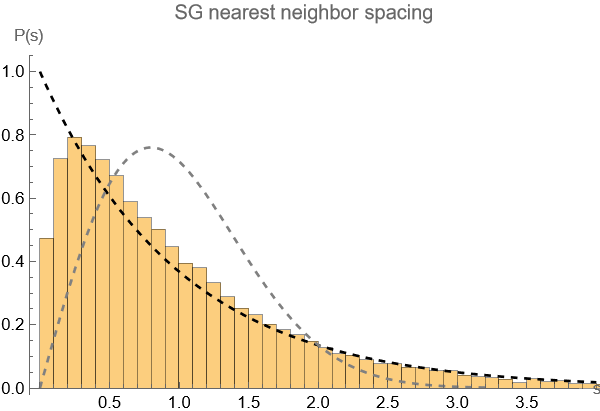}
         \caption{}
         \label{fig:SG-NN-spacing}
     \end{subfigure}
   \begin{subfigure}{\columnwidth}
         \centering
        \includegraphics[width=\linewidth]{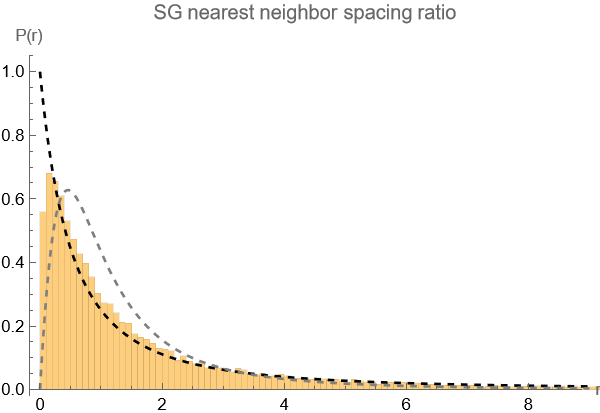}
         \caption{}
         \label{fig:SG-NN-spacing-ratio}
     \end{subfigure}
            \caption{Level spacing and spacing ratios statistics from the SG data. Fig. \ref{fig:SG-NN-spacing} shows the consecutive level spacing distribution, where the black dashed line is the RMT prediction for a Poisson distribution (\ref{eq:Poisson-distribution}) and the grey dashed line is the GOE WD distribution (\ref{eq:WD-distribution}). Fig. \ref{fig:SG-NN-spacing-ratio} shows the consecutive level spacing ratio distribution, with the black dashed line being the Poisson ensemble prediction for level spacing ratios (\ref{eq:NN-spacing-ratio-distribution-Poisson}) and the grey dashed line being the GOE level spacing ratio from RMT, (\ref{eq:P(r,beta,k)}).}
        \label{fig:SG-NN}
\end{figure}

From Fig. \ref{fig:SG-NN} we can see in both the level spacing and level spacing ratios that the data does not fit the Poisson distributions neatly - there is a clear level repulsion for $s,r\rightarrow 0$, which should not exist for pure integrable models, and a following 'excess' of levels for $s,r$ a bit larger. The tails of both the spacing and the ratios do seem to fit the Poisson prediction. Both cases are far from the GOE distributions which are denoted by the gray curves,in the language of a mixing parameter it seems closer to the integrable value. SG is an integrable model, so these results are attributed to the numerical truncation in the calculation of the spectrum. We also calculate the mean spacing ratio $\langle \tilde{r} \rangle=0.4361$, which is greater than the RMT predicted $\langle \tilde{r} \rangle_{PE}=0.3863$.\par
Taking SG data with $mL<1$ yields better results. Using the unfolded data of SG with $mL=0.01$, which corresponds to taking a higher cutoff energy, gives results that fit the RMT predictions much better, with $\langle \tilde{r} \rangle_{mL=0.01}=0.3819$. Thus we use $mL=0.01$ for all further calculations of the SG data.

\section{Distributions of non-adjacent  spacings and their ratios}\label{Dona}
\begin{figure}[tbp]
     \centering
   \begin{subfigure}{\columnwidth} 
         \centering
        \includegraphics[width=\linewidth]{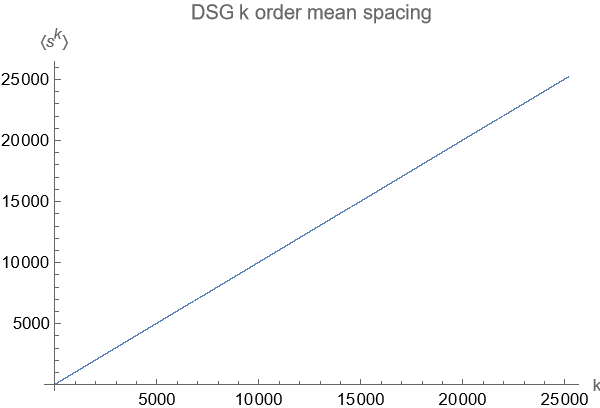}
         \caption{}
         \label{fig:DSG-mean-k-spacing}
     \end{subfigure}
     \medskip
   \begin{subfigure}{\columnwidth}
         \centering
        \includegraphics[width=\linewidth]{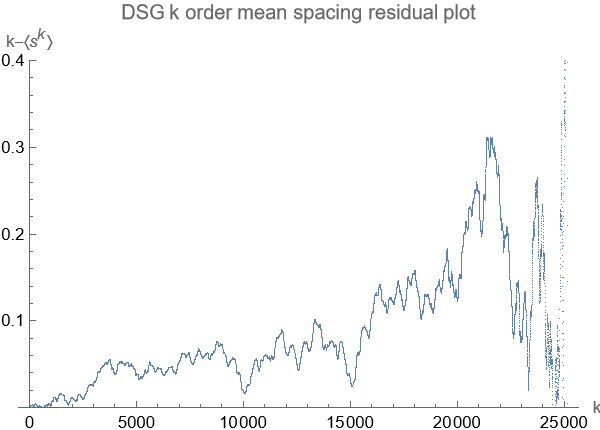}
         \caption{}
         \label{fig:DSG-mean-k-spacing-residuals}
     \end{subfigure}
        \medskip
        \begin{subfigure}{\columnwidth} 
         \centering
        \includegraphics[width=\linewidth]{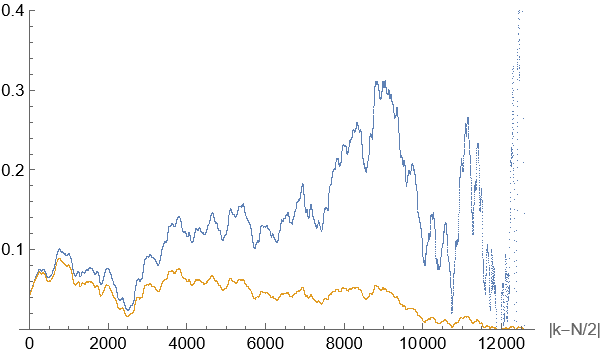}
         \caption{}
         \label{fig:DSG-mean-k-spacing-residuals-reflection}
     \end{subfigure}
        \medskip
            \caption{(\ref{fig:DSG-mean-k-spacing}) The mean $k$ spacing ratio $\langle r^k \rangle$ from the unfolded DSG high energy levels. It shows a linear behavior with a slope of 1. The deviation from the predicted linear behavior is seen in Fig. (\ref{fig:DSG-mean-k-spacing-residuals}). Fig. (\ref{fig:DSG-mean-k-spacing-residuals-reflection}) shows the similarity of the residuals under reflection around $k=N/2$.}
        \label{fig:DSG-high-k-mean-spacing}
\end{figure}

We also measure the mean $k$-spacing and spacing ratio, as elaborated in section \ref{higher-orders}. Fig. (\ref{fig:DSG-mean-k-spacing}) shows that the mean $k$ behaves linearly in $k$ with a slope of 1. We suggest an explanation for this behavior in section \ref{Probabilistic-approach}. The deviations of $\langle s^k \rangle$ from linearity are very small, as can be seen from Fig. (\ref{fig:DSG-mean-k-spacing-residuals}).

The pattern in the residual plot looks random at first, but in closer inspection there seems to be a reflection symmetry around the value of $k \approx 12,500$, which happens to be $k=N/2$. This symmetry is masked by the distortion of the values for $k>N/2$, but that distortion is due to the shrinking size of the samples of spacings, which is a linear function in $k$ as well. $$N_{samples}(k)=N-k-1$$ To see the symmetry and the distortion, we take the residual plot of Fig. (\ref{fig:DSG-mean-k-spacing-residuals}) and reflect the plot around $k=N/2$  
 in Fig. (\ref{fig:DSG-mean-k-spacing-residuals-reflection}) such that the x axis is now the distance from the reflection point, $|k-N/2|$ and the orange curve in the reflected part of the plot. The blue curve is the part unchanged where $k>N/2$. We find that the two curves indeed correspond, though the distortion shifts their values apart.

\begin{figure}[tbp]
    \centering
    \includegraphics[width=\columnwidth]{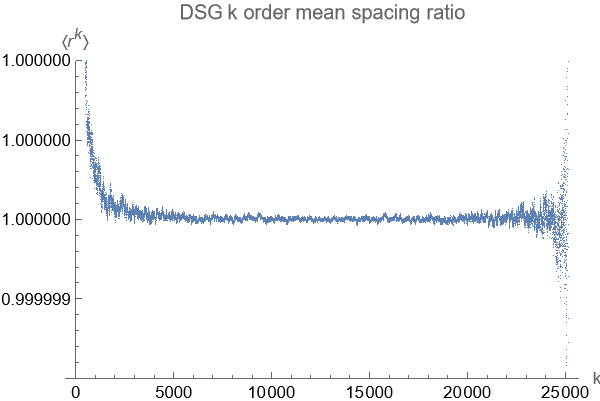}
    \caption{Mean $k$-spacing ratio, plotted against $k$.}
    \label{fig:DSG-high-k-mean-spacing-ratio}
\end{figure}

We also compute the mean $k$ spacing ratios, seen in Fig. (\ref{fig:DSG-high-k-mean-spacing-ratio}). Since there is no prediction to compare the results with, we'll note that the mean is predominantly 1, with fluctuations 7 orders of magnitude smaller, as can be seen from the main part of Fig. (\ref{fig:DSG-high-k-mean-spacing-ratio}). Again, though it is harder to see in this case, we find the reflection symmetry around $k=N/2$, and a distortion of the mean values around the value of 1, which grows with $k$ as the sample size shrinks, though the distortion seems to be prominent for $k$'s much larger than in the spacing case.

For small $k$, we can check the validity of the results with the conjecture in \cite{HarishniTekur} where the probability distribution of the $k$ spacing ratios is given by (\ref{eq:P(r,beta,k)})

\begin{equation}
\label{eq:P(r,beta,k)}
P(r,\beta') = \frac{1}{N_\beta'} \frac{(1+r)^{\beta'}}{(1+r+r^2)^{1+3\beta'/2}}
\end{equation}

with $$\beta'=\frac{k(k+1)}{2}\beta + (k-1)$$ and compare the DSG results with RMT results and (\ref{eq:P(r,beta,k)}).\par

\begin{figure} [tbp]
     \centering
   \begin{subfigure}{\columnwidth} 
         \centering
     \includegraphics[width=\linewidth]{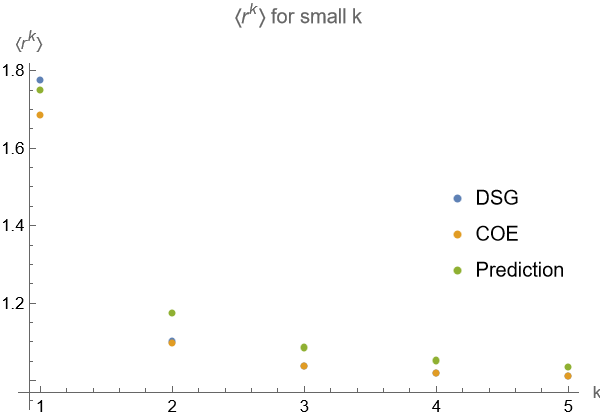}
         \caption{}
         \label{fig:DSG-mean-k-spacing-ratio-comparison}
     \end{subfigure}
     
   \begin{subfigure}{\columnwidth}
         \centering
          \includegraphics[width=\linewidth]{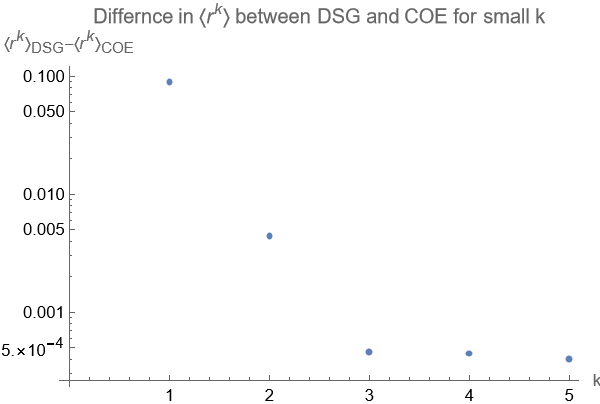}
         \caption{}
         \label{fig:DSG-mean-k-spacing-ratio-DSG-COE}
     \end{subfigure}
     
            \caption{(\ref{fig:DSG-mean-k-spacing-ratio-comparison}) $\langle r^k \rangle$ for the first 5 spacing ratios calculated from the DSG unfolded data, from the Circular Orthogonal Ensmeble (COE) with $N=5000$ and from the analytical expression given in (\ref{eq:P(r,beta,k)}). Aside from $k=1$, the DSG and COE results are indistinguishable. To get a better understanding of the comparison of DSG and COE, in (\ref{fig:DSG-mean-k-spacing-ratio-DSG-COE}) we plot in a log scale the difference between the mean ratios for each k.}
        \label{fig:DSG-mean-k-spacing-ratio}
\end{figure}

The DSG and RMT results match even for small k, as can be seen in Fig. (\ref{fig:DSG-mean-k-spacing-ratio-DSG-COE}) where the difference is of order $10^{-4}$ already for $k=3$.

For higher order spacings and spacing ratios of integrable systems, there is an expression due to \cite{ABGVV} for the $k$-spacing ratio distribution

\begin{equation}
\label{eq:P(r,k)-integrable}
 P_{k}(r) =
  \begin{cases}
 \frac{r^{k}(1+k+kr)}{(1+r)^2} & \text{for } r \leq 1 \\\\
 \frac{k+r(k+1)}{r^{k+1}(1+r)^2} & \text{for } r > 1
  \end{cases}
\end{equation}

From which we can derive the mean $k$-spacing ratio

\begin{multline}
\langle r_k \rangle
= \frac{1}{2} \left[ 2+H\left(\frac{k}{2}\right)-H\left(\frac{k+1}{2}\right)-H\left(\frac{k-2}{2}\right) \right. \\
\left. + H\left(\frac{k-1}{2}\right) \right]
\label{eq:Mean-r_k-for-integrable-systems}
\end{multline}

\begin{figure} [tbp]
     \centering
   \begin{subfigure}{\columnwidth} 
         \centering
          \includegraphics[width=\linewidth]{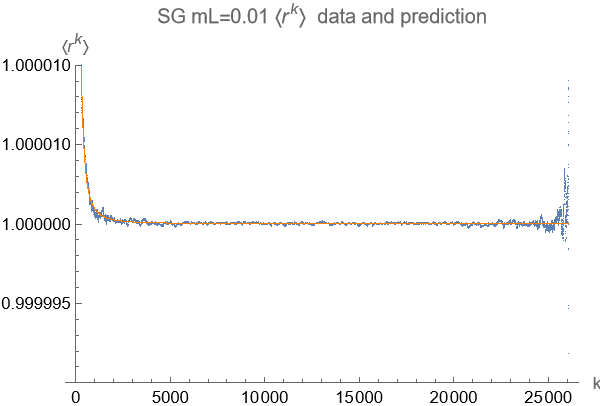}
         \caption{}
         \label{fig:SG-mean-k-ratio}
     \end{subfigure}
     
   \begin{subfigure}{\columnwidth}
         \centering
          \includegraphics[width=\linewidth]{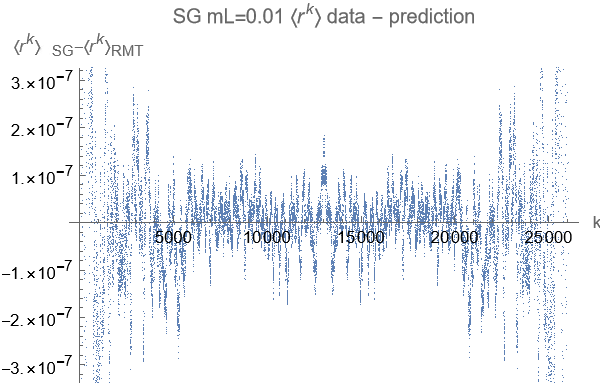}
         \caption{}
         \label{fig:SG-mean-k-ratio-res}
     \end{subfigure}
     
            \caption{\ref{fig:SG-mean-k-ratio} shows $\langle r_k \rangle$ calculated from the SG data (blue) together with the prediction (\ref{eq:Mean-r_k-for-integrable-systems-Asymptotically}) (orange), which is easier to compute than (\ref{eq:Mean-r_k-for-integrable-systems}) for large $k$. \ref{fig:SG-mean-k-ratio-res} is the residual plot, the difference between the calculated means and the prediction. Notice the small scale of $10^{-7}$ of the fluctuations of the data, and the appearance of the $k=N/2$ reflection symmetry.}
\end{figure}

Where $H(k)$ are the harmonic numbers, defined as $H(k)=\sum_{n=1}^{k} \frac{1}{n}$ for a finite integer $k$. Asymptotically, (\ref{eq:Mean-r_k-for-integrable-systems}) can be written as

\begin{equation}
\label{eq:Mean-r_k-for-integrable-systems-Asymptotically}
\langle r_k \rangle\approx 1+\frac{1}{2}\log\left(\frac{k(k-1)}{(k+1)(k-2)}\right)
\end{equation}

where it is clearer that $\langle r_k \rangle$ converges to 1 as $k\rightarrow\infty$. We can compare (\ref{eq:Mean-r_k-for-integrable-systems}) with the mean $k$-spacing ratio calculated from the data, as seen in Fig. (\ref{fig:SG-mean-k-ratio}). The prediction agrees with the calculation from the data, and taking the difference of the two shows that the $k=N/2$ reflection symmetry seems to hold for integrable cases as well.\par

\begin{figure} [tbp]
     \centering
   \begin{subfigure}{\columnwidth}
         \centering
        \includegraphics[width=\linewidth]{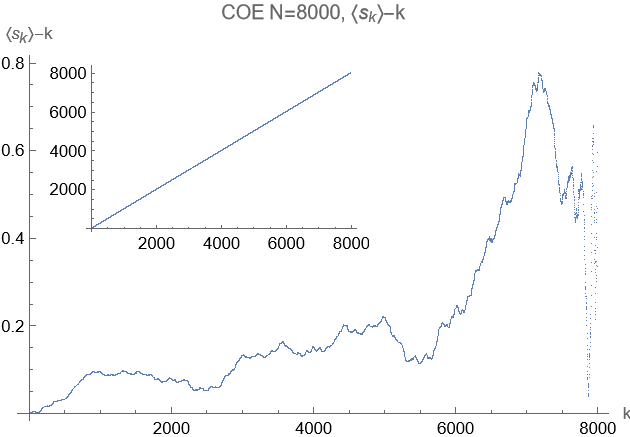}
         \caption{}
         \label{fig:COE-N=8000-with-mean-k-inset}
     \end{subfigure}
   \begin{subfigure}{\columnwidth}
         \centering
        \includegraphics[width=\linewidth]{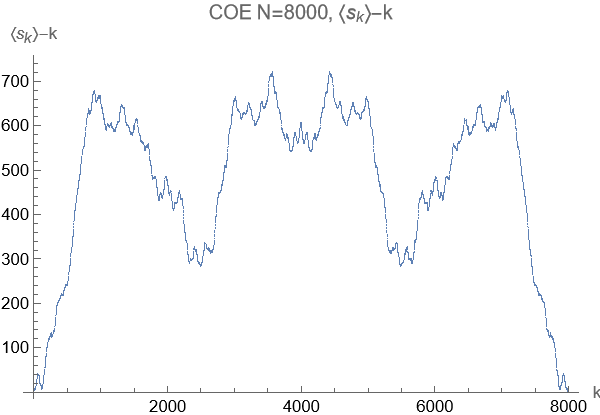}
         \caption{}
         \label{fig:COE-N=8000-normalized-deviation-from-k}
     \end{subfigure}
     
            \caption{data taken from COE with N=8000. The inset in Fig. \ref{fig:COE-N=8000-with-mean-k-inset} shows the mean $k$-spacing of the data as a function of $k$, which in this scale is indistinguishable from (\ref{eq:k-spacing-mean}). The rest of Fig. \ref{fig:COE-N=8000-with-mean-k-inset} is the deviation of the mean $k$-spacing from (\ref{eq:k-spacing-mean}). Fig. \ref{fig:COE-N=8000-normalized-deviation-from-k} is the same as Fig. \ref{fig:COE-N=8000-with-mean-k-inset} but multiplied by a factor of $N_{sample}$ to compensate for an apparent $1/N_{sample}$ distortion. The result show a clear reflection symmetry around the $k=N/2$ value.}
        \label{fig:total-COE-N=8000}
\end{figure}

In conclusion, we have seen that the mean $k$-spacing and spacing ratios show fluctuation around the predictions. For example $$\langle s_k \rangle = k + \langle s_k \rangle_{fl} $$ where $\langle s_k \rangle_{fl}$ is of $O(1)$ regardless of the size of $k$ and obey a reflection symmetry around $k=N/2$, as can be seen in Fig. \ref{fig:DSG-mean-k-spacing-residuals-reflection}, \ref{fig:SG-mean-k-ratio-res}, \ref{fig:COE-N=8000-normalized-deviation-from-k}. We notice that it is present for means on the whole spectrum, as in the COE example of section \ref{Probabilistic-approach}, and for means on partial parts of the spectrum, as in the SG and DSG calculations, and that it is present in integrable (SG) and non-integrable (DSG) cases, so it seems to be robust phenomena that is unrelated to the data itself. We also notice that such a symmetry is not present in the $k$-standard deviation, only in the fluctuations from the mean. We suggest that this symmetry is related to the autocorrelation structure of the mean: given a discrete spectrum $\{e_i\}$ of size $N$, the mean $k$-spacing is explicitly $$\langle s_k \rangle=\frac{1}{N-k}\sum_{n=1}^{N-k} (e_{n+k}-e_n)$$ And in the continuum limit we add a delta function to fix the distance $s$ between the two levels $e,e'$ 

\begin{equation}
\begin{split}
\
\frac{1}{N-s}\int de\int de'\rho(e)\rho(e')(e-e')\delta(e-e'-s)\\
=\frac{1}{N-s}\int de\rho(e)\rho(e-s)s\
=\frac{s}{N-s}\int de\rho(e)\rho(e-s)
\end{split}
\label{eq:autocorrelation}
\end{equation}

Where in the last term, the integral is the autocorrelation of $\rho(e)$ with lag $s$, $(\rho\star\rho)(s)$, which has a reflection symmetry around $s=N/2$. In the last calculation we should not confuse the continuous lag $s$, which is analogous to the discrete spacing $k$, with the symbol $s$ for spacing in the brackets of $\langle s_k \rangle$. 

\subsection{Probabilistic approach to higher order level spacings} \label{Probabilistic-approach}

In the light of section \ref{higher-orders}, we suggest a naive approach to the linear behavior of the mean $k$-spacing: given an energy level $E_n$, WD statistics suggests that the consecutive energy level $E_{n+1}$ can be constructed as $E_{n+1}=E_n + S$ where $S\sim P(s)$ is a random variable drawn from the NN spacing probability distribution $P(s)$. This means that for the $k$ separated energy levels, the $k$-spacing is

\begin{equation}
\label{eq:k-spacings}
s_{n,k}=E_{n+k}-E_{n}=\sum_i^k S_i
\end{equation}

which is the convolution of $k$ NN PDFs. It follows that  the mean is 

\begin{equation}
\label{eq:k-spacing-mean}
\langle s_k \rangle =\langle\sum_i^k S_i\rangle = k \langle s \rangle
\end{equation}

and for normalized NN distributions, where $\langle s \rangle=1$, we get that $\langle s_k \rangle=k$.
We can also see that for large $k$, using the central limit theorem

\begin{equation}
\label{eq:k-spacing-central-limit-theorem}
\lim_{k\to\infty} s_{n,k}\sim N\left(k,\sigma_{NN}\sqrt{k}\right)
\end{equation}

Where $N$ is a normal probability distribution with mean $\mu=k$ and standard deviation $\sigma=\sigma_{NN}\sqrt{k}$. The contribution of the choice of NN distribution comes only in $\sigma_{NN}$ when $k$ is large.
Note that like other approaches that are based on local statistics, we assume a constant mean level density, either by construction (as in a normalized circular ensemble) or by unfolding.\par
We check this approach numerically using randomly generated eigenvalues of the COE with $N=5000$ and $N=8000$. The mean does seem to behave linearly in $k$ as suggested by (\ref{eq:k-spacing-mean}). When looking at the deviation from linearity, we find the deviations to be of order 1, even for large $k$. If we multiply the deviations in the mean with a factor of the size sample for each $k$, which is $N_{sample}=N-k$, we cancel a $1/N_{sample}$ factor and get a picture that shows a reflection symmetry around $k=N/2$.

On the other hand we can calculate the expression $P_{k}(s)$ for $k=2$ analytically by convolving two WD distributions with $\beta=1$. We get 

\begin{multline}
P_{k=2}(s)
=\frac{\pi}{16}\exp\!\left(-\frac{\pi s^2}{4}\right)\Bigl[
4s \\
+\sqrt{2}\exp\!\left(\frac{\pi s^2}{8}\right)
\left(\pi s^2 - 4\right)\,
\operatorname{erf}\!\left(\frac{1}{2}\sqrt{\frac{\pi}{2}}\,s\right)
\Bigr]
\label{eq:k=2-convolution}
\end{multline}

\begin{figure}[tbp]
    \centering
    \includegraphics[width=\linewidth]{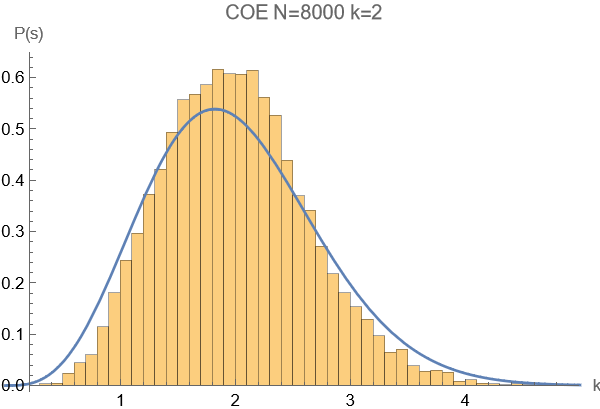}
    \caption{COE N=8000 k=2 spacing distribution (orange bar) with prediction from the convolution of two WD $\beta=1$ probability distributions, (\ref{eq:k=2-convolution}, blue curve). The spacing data, as opposed to (\ref{eq:k=2-convolution}), has apparent surplus of spacings around the mean, at the expense of less spacing in the higher and lower values. This is also indicated by a smaller standard deviation of the data, which is around $0.633$, as opposed to the prediction, which is $\sqrt{\frac{8}{\pi}-2}\sim0.739$.}
    \label{fig:COE-N=8000-k=2-with-P_2(s)}
\end{figure}

Where $Erf(x)=\frac{2}{\sqrt{\pi}}\int_0^x \exp{-t^2}dt$ is the error function. When comparing (\ref{eq:k=2-convolution}) with the numerical $k=2$ spacing distribution (Fig. \ref{fig:COE-N=8000-k=2-with-P_2(s)}) we get the mean is $2$ for both, but the numerical data has a smaller standard deviation.

\begin{figure} [tbp]
     \centering
   \begin{subfigure}{\columnwidth}
         \centering
    \includegraphics[width=\linewidth]{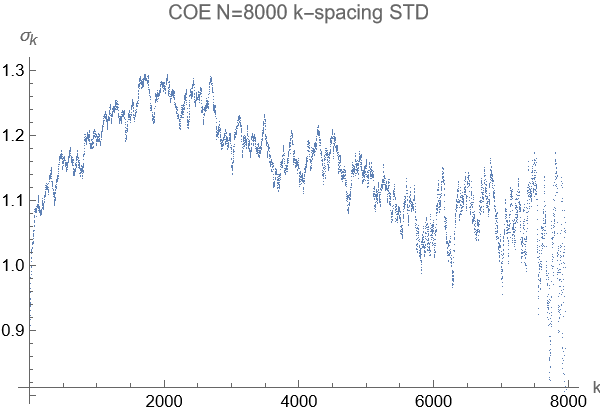}
         \caption{}
         \label{fig:COE-N=8000-mean-k-spacing-STD}
     \end{subfigure}
   \begin{subfigure}{\columnwidth}
         \centering
        \includegraphics[width=\linewidth]{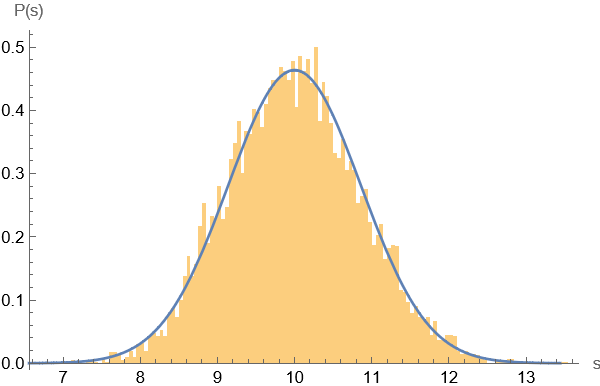}
         \caption{}
         \label{fig:COE-N=8000-k=10-with-Gaussian}
     \end{subfigure}
    \begin{subfigure}{\columnwidth}
         \centering
        \includegraphics[width=\linewidth]{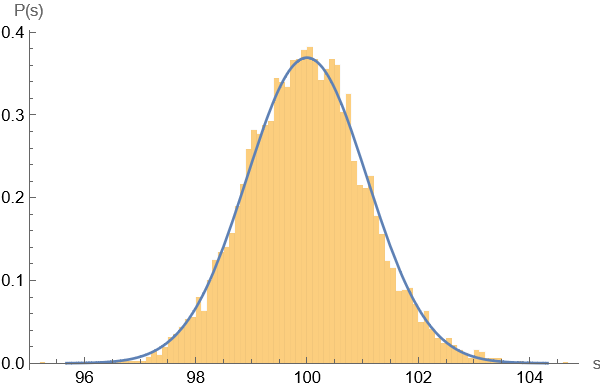}
         \caption{}
         \label{fig:COE-N=8000-k=100-with-Gaussian}
     \end{subfigure}
    \begin{subfigure}{\columnwidth}
         \centering
        \includegraphics[width=\linewidth]{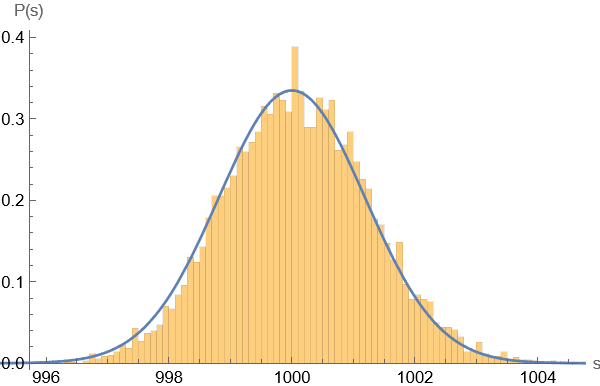}
         \caption{}
         \label{fig:COE-N=8000-k=1000-with-Gaussian}
     \end{subfigure}
            \caption{Fig. \ref{fig:COE-N=8000-mean-k-spacing-STD} shows the standard deviation of k-spacing distribution calculated from COE with $N=8000$. Note that it does not follow the $\sim\sqrt{k}$ prediction of (\ref{eq:k-spacing-central-limit-theorem}) .Fig. \ref{fig:COE-N=8000-k=10-with-Gaussian}-\ref{fig:COE-N=8000-k=1000-with-Gaussian} shows different $k$-spacing distributions from the COE data, together with normal distributions with mean k and standard deviations from the corresponding data, which are seen for all $k$ in Fig. \ref{fig:COE-N=8000-mean-k-spacing-STD} (blue curves). In Fig. \ref{fig:COE-N=8000-k=1000-with-Gaussian} the data seem to be bulged slightly to the right, since the mean is actually $1000.09$, as can be seen from  Fig. \ref{fig:COE-N=8000-with-mean-k-inset}.}
        \label{fig:COE-N=8000}
\end{figure}

When we go to larger $k$, we see that even for $k$ as small as 4 or 5, the central limit theorem is a good enough approximation and the numerical distribution fits nicely with a normal distribution of mean $k$. The standard deviation, however, differs from $\sigma_{NN}\sqrt{k}$ and remains in same order of magnitude as $\sigma_{NN}$ for all $k$, as can be seen in Fig. \ref{fig:COE-N=8000-mean-k-spacing-STD}. We emphasize these results by showing how the $k$-spacing distributions of the data fit nicely with normal distributions with mean $k$ and standard deviations taken from the data, $N(k,\sigma_{k,data})$ (Fig. \ref{fig:COE-N=8000-k=10-with-Gaussian}-\ref{fig:COE-N=8000-k=1000-with-Gaussian}). We suggest that the discrepancy between the naive probabilistic approach and the data comes from the assumption that the random variables in (\ref{eq:k-spacings}) are independent, whereas ergodicity implies that the energy levels are very dependent on each other, even if they are separated by $k$ other energy levels.

\section {Pair correlation}\label{Pc}

We now turn to long range correlations using higher order spacings and calculate the pair correlation. Using the unfolded data of DSG we calculate the level spacing for all orders, and show the normalized probability distribution of the short spacings (around order of 1. Note the unfolding rescaled the energy levels such that the mean consecutive level spacing is 1). Comparing that with the 2-point cluster function calculated from the GOE, which reads \cite{Mehta}

\begin{equation}
\label{eq:Y_GUE}
Y_{GOE}(r) = \left( \frac{1}{2}-\int_0^{r}S(r')dr' \right) \left( 
\frac{dS}{dr} \right) + S(r)^2
\end{equation}

Where $$S(r)=\frac{\sin(\pi r)}{\pi r}$$ and $r$ is the spacing between any two energy levels, we find a high level of  agreement between the data and the prediction.

\begin{figure} [tbp]
    \centering
    \includegraphics[width=8cm]{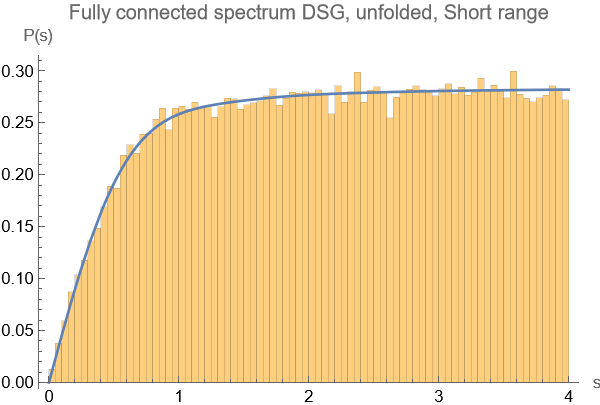}
    \caption{DSG all order (not only nearest neighbor) spacings 
 (orange bars) with the predicted $1-Y_{GOE}$ (blue curve), normalized to the range shown. $Y_{GOE}$ is given by (\ref{eq:Y_GUE}).}
    \label{fig:DSG-pair-correlation}
\end{figure}

Following the discussion in (\ref{eq:2-point-correlation-GUE}), the cluster function for integrable systems is 0 so the correlation function, after normalization, is 1. The calculation of the pair correlations shows indeed a 'plateau' when looking at the short range of spacings, and at the full range it has the triangular shape due to the pair correlation being the autocorrelation of the level densities, which are constant at scales larger than the scale in which the fluctuations average to that constant, which is the same order as the correlation length - the length from which level repulsion becomes significant when there is one. The average level density is of course constant due to the unfolding.

\section{Spectral form factor}\label{Sff}

\begin{figure} [tbp]
     \centering
   \begin{subfigure}{\columnwidth}
         \centering
          \includegraphics[width=\linewidth]{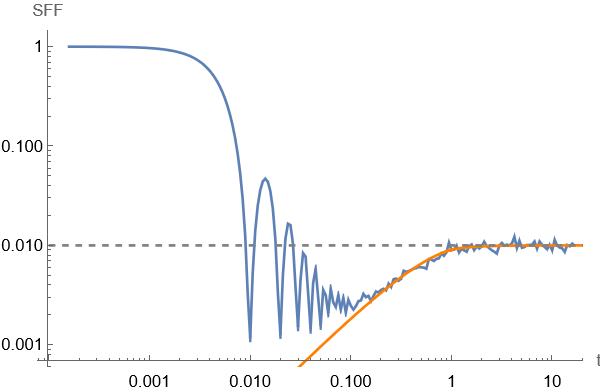}
         \caption{DSG SFF}
         \label{fig:DSG-SFF}
     \end{subfigure}
     
   \begin{subfigure}{\columnwidth}
         \centering
          \includegraphics[width=\linewidth]{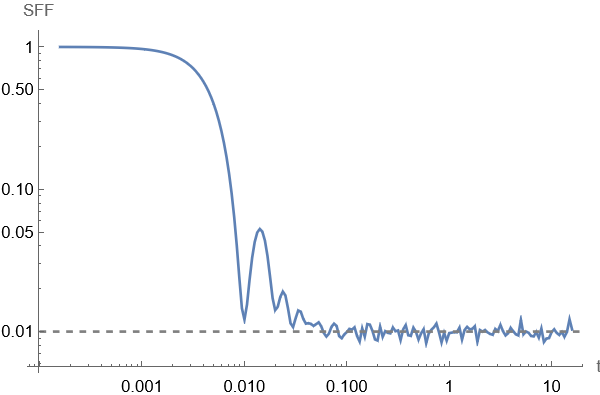}
         \caption{SG SFF}
         \label{fig:SG-SFF}
     \end{subfigure}
     
            \caption{average SFF of DSG and SG as a function of time $t$ (blue curves). The orange curve in (\ref{fig:DSG-SFF}) is the GOE SFF given in (\ref{eq:SFF-connected-GOE}), with $t\rightarrow t/2\pi$. The dashed line in both (\ref{fig:DSG-SFF}) and (\ref{fig:SG-SFF}) is $1/N$, which is the asymptotic value of the averaged SFF for both chaotic and integrable cases.}
        \label{fig:SFF-DSG-and-SG}
\end{figure}

We proceed and calculate the spectral form factor for the unfolded spectra of SG with $mL=0.01$ and DSG with $mL=1$. To do so we choose from the higher end of the spectrum energy windows of size $N=100$ and use (\ref{eq:discrete-SFF}) to calculate the SFF as a function of time. We repeat that calculation 250 times for different energy windows which are mutually disjoint, and finally take the average of all those calculations. The result can be seen in fig. \ref{fig:SFF-DSG-and-SG}. The DSG shows structure characteristic of chaotic spectra, where there is initially an oscillating descent that goes to values smaller than $1/N$, then a ramp which should be linear for systems with time reversal symmetry that correspond to the GOE, as in (\ref{eq:SFF-connected-GOE}) when $t\leq1$, and finally a plateau around the value of $1/N$ when $t\geq1$. The region where the value of the SFF is smaller than $1/N$ is called the correlation hole. On the other hand, the SFF from SG exhibits no correlation hole, and the SFF is made of an oscillating descent that stops at the plateau at the value of around $1/N$. This is characteristic behavior of SFF of integrable systems, where the energy levels are uncorrelated.

\begin{figure} [tbp]
    \centering
    \includegraphics[width=8cm]{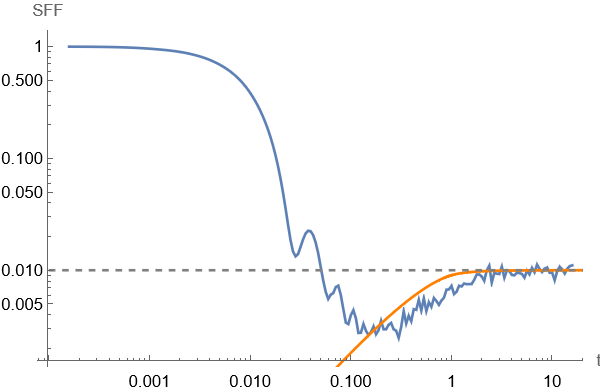}
    \caption{SFF from the DSG spectrum which normalized but not unfolded, together with the GOE connected SFF (orange curve).}
    \label{fig:SG-SFF-unfolded}
\end{figure}

We compare this to the SFF of the DSG spectrum that was not unfolded, only normalized such that the mean NN spacing is 1 (see fig. \ref{fig:SG-SFF-unfolded}). Though the correlation hole is still visible, the SFF from unfolded spectrum is a better fit to the GOE curve than the one from the regular spectrum. We note that even without the unfolding, the SFF is not far from the GOE curve, since the change of the average mean level density is small in the energy windows chosen.

\section{Spectral Rigidity}\label{SR}

Like the SFF and pair correlation, the spectral rigidity (also known as $\Delta_3$ statistics) is a measure of the long-range correlation between the energy levels. It shows distinct behavior for integrable systems, where it is linear in $L$, as opposed to chaotic systems where it goes like $\sim\ln(2\pi L)$. 
Using (\ref{eq:spectral-rigidity}), we calculate the spectral rigidity of the unfolded SG and DSG data, and compare them to the RMT expressions in (\ref{eq:spectral-rigidity-asymptotics}).
In accordance with previous calculations, we choose $mL=1$ for DSG and $mL=0.01$ for SG, and take the high range part of the spectrum.

\begin{figure}[tbp]
     \centering
   \begin{subfigure}{\columnwidth}
         \centering
     \includegraphics[width=\linewidth]{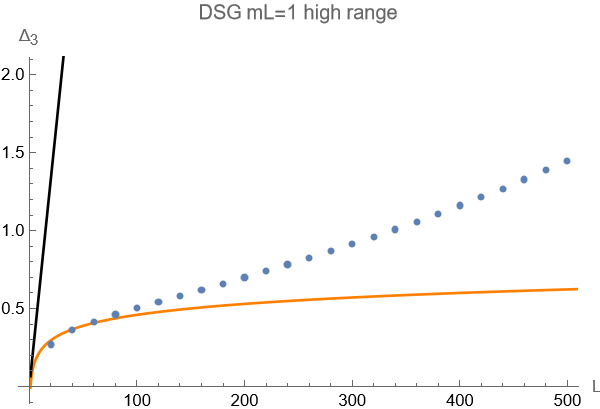}
         \caption{DSG spectral rigidity}
         \label{fig:DSG-spectral-rigidity}
     \end{subfigure}
     
   \begin{subfigure}{\columnwidth}
         \centering
     \includegraphics[width=\linewidth]{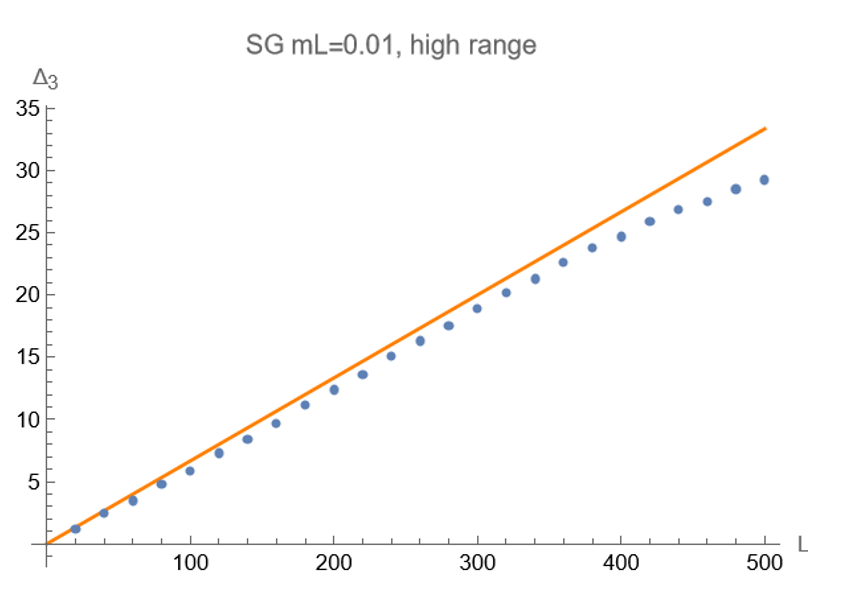}
         \caption{SG spectral rigidity}
         \label{fig:SG-spectral-rigidity}
     \end{subfigure}
          
            \caption{The numerically calculated spectral rigidity of SG and DSG, calculated from the high range of the unfolded data, seen as blue dots. In \ref{fig:DSG-spectral-rigidity} The oragne curve is $\Delta_{3,1}$ in (\ref{eq:spectral-rigidity-asymptotics}) and the black curve is $\Delta_{3,0}$. The numerical calculations follow $\Delta_{3,1}$ until $L\sim70$, where they detach from the curve. The orange curve in \ref{fig:SG-spectral-rigidity} is $\Delta_{3,0}$}
        \label{fig:spectral-rigidity}
\end{figure}

We find that SG fits well with $\Delta_{3,0}=\frac{1}{15}L$ for $L$ up to $500$. For DSG,  fig. (\ref{fig:DSG-spectral-rigidity}) shows that for small $L$ the data fits with analytical expression (the oragne curve), but around $L\sim70$ it diverges towards the integrable analytical curve (the black curve). This suggests that even though for the local measures, the DSG data seems to fit well with GOE, there's a limit to the range where the energy levels are correlated, and above that range Poisson statistics dominate.
This behavior is also produced in \cite{PhysRevE:52:148}, where the authors use a random matrix model that smoothly interpolates between between Poisson-like and GOE statistics, using a mixing parameter $\Delta\in[0,1]$. For mixed spectra, the spectral rigidity starts on the GOE curve and diverges around a value $L_{max}$, which is a function of both the mixing parameter $\Delta$ and the size of the matrices $N$, such that for $N\rightarrow\infty$ and $\Delta>0$, $L_{max}\rightarrow\infty$ and the GOE statistics $\Delta_{3,1}$ are recovered. It is also seen in the framework of many-body localization transition (MBLT) \cite{PhysRevB.93.041424}, where the Thouless energy, that relates to the fractal dimension of the wave function near the mobility edge, sets the scale below which the statistics of the energy level obey RMT, and above which show Poisson statistics. The Thouless time $t_{Th}$ relates to the Thouless energy $E_{Th}$ via $t_{Th}=\hbar/E_{Th}$, and can be calculated using the dipping time $t_p$ in the SFF and the Heisenberg time $t_H=2\pi/\langle s \rangle$ by $t_{Th}=t_Ht_p$ \cite{Sierant_2020}, where $\langle s \rangle$ is the mean NN spacing before the unfolding. Using the calculation seen in fig. (\ref{fig:DSG-SFF}) we get $t_{Th}\sim44,500$.
In any case, one cannot rule out the effect of the finiteness of the energy levels on the correlation length.
See section IV.4 in \cite{Weidenmuller} for a further discussion.

\section{Conclusions and future directions} \label{conclusion}

We have computed various measures of characteristic quantum chaotic behavior on numerically calculated spectra of SG and DSG.
\begin{itemize}
\item    
We have found that the DSG data when $mL=1$ matches the behavior of the GOE in the NN spacing distribution and NN spacing ratio distributions. We found that mean value of the nearest neighbor spacing ratio deviate by less than $1\%$ from the value of the GOE.
For the SG model with $mL=1$ the correspondence to the Poisson distributions  is not as good as for the DSG.. Using the unfolded data with $mL = 0.01$,  gives results that fit the RMT predictions much better.  
\item
We computed the means of higher order level spacing distributions. The mean $k$-spacing $\langle s_k\rangle$ shows a clear linear behavior with a slope of 1, and the mean $k$-spacing ratio $\langle r_k \rangle$ matches the analytical prediction \ref{eq:Mean-r_k-for-integrable-systems-Asymptotically}from the unfolded DSG high energy levels. It shows
a linear behavior with a slope of 1.
\item
We have suggested a probabilistic interpretation to the $k$-spacing distribution which matches the Gaussian like distributions that are obtained for large $k$, and the linear relation of mean $k$-spacing $\langle s_k \rangle$ and $k$. This approach falls short when it comes to predictions of the standard deviation in $k$-spacing distributions, and it is less accurate in the specific distributions of small $k$'s, both cases due to the assumption that the distribution of each energy level is uncorrelated to the other, which is true only for the Poisson Ensemble. We would like to measure the deviation of these distributions from Gaussianity in order to quantify this dependence.
\item 
We have computed several long-range correlation measures. The pair correlation shows good agreement with the RMT prediction (\ref{eq:Y_GUE}). The SFF also seems to agree with the predicted connected part (\ref{eq:SFF-connected-GOE}). As for the spectral rigidity - SG matches the prediction well for all $L$, but in DSG we see that the numerically calculated spectral rigidity matches the prediction (\ref{eq:spectral-rigidity-asymptotics}) only up to a certain value of $L$, for which we have also computed the corresponding Thouless time. This indicates that the GOE behavior of DSG is scale dependent, and beyond a certain energy scale the energy levels are uncorrelated.

\end{itemize}
\subsection*{Future directions}

Other directions we would like to pursue in future research are:

\subsubsection*{quantum chaos characteristics}

There are several more measures for quantum chaoticity that where not studied  in this work and we would like to implement in the future:

\begin{itemize}
\item \textit{Spectral rigidity}- We intend to further study the discrepancy between the expected spectral rigidity as a function of $L$ and  the observed result.
\item \textit{Lyapunov exponents (OTOCs)} - These where mentioned in section 1. We would like to explore further the connection to RMT and its values for QFT.
\item \textit{Krylov complexity} - Also known as $K-complexity$, an increasingly popular measure in recent times, specifically in describing the spread of local operators in Hilbert space as they evolve in time. The late time Krylov complexity of a system is different in a distinguished way for integrable and non-integrable systems.\cite{Rabinovici}\cite{Bhath:25}.
\end{itemize}

\subsubsection*{Transient quantum chaos and scattering}

Though the study of quantum chaos usually relies on the energy spectrum or eigenmodes of models, one can also study the quantum analogs of classically transient chaos, i.e. the chaos of scattering \cite{Gross:2021gsj},\cite{Rosenhausstrings},\cite{Bianchi:2022mhs}, \cite{Cobistrings},\cite{CobScFF}. A canonical classical example is the erratic and self similar behaviors of 'pinballs' - the scattering of a particle of a potential of 3 hard disks. The chaotic behavior is seen by plotting the final deflective angle of the particle as a function of the impact parameter \cite{Rosenhaus}, or in the transient time the particle spends inside the potential before shooting off . In quantum systems this behavior is described by the S-matrix, and the chaotic behavior could be described by the decay time or deflective angle of a highly excited state. This approach is more suitable for QFT since the S-matrix is a more natural observable than the energy levels.

\subsubsection*{Applications to other QFT}

We would like to explore chaoticity using the measures mentioned above for more $(1+1)$D QFT, by applying the TCSA \cite{TCSA} or light-cone conformal truncation \cite{LightconeTrunc} methods to compute the spectrum of integrable models and non-integrable perturbations thereof by controlling some order parameter. Some possible directions are:

\begin{itemize}
\item \textit{CFT ensembles} - A recent work \cite{Sonner} proposes a CFT analog to RMT, where each member of the ensemble is an approximate data set of a CFT.
\item \textit{Wess-Zumino Witten Model} - Well studied case of $(1+1)$D CFT with several sectors which could show chaotic behavior. \cite{WZW}.
\item \textit{2D QCD} - At $N\rightarrow\infty$ 't Hooft limit  where the model is integrable and $1/N$ perturbations may introduce chaotic behavior \cite{tHooft}.
\item \textit{Bosonized 2D QCD} - multiflavored bosonized 2D QCD in the baryonic sector \cite{Bosonization}.
\item higher dimensional CFT such as the 4D $\mathcal{N}=4$ SYM in the large N limit, by studying the spectrum from spin chains \cite{SYM}, and non-integrable $1/N$ perturbations thereof.
\item higher dimensional QFTs in general, such as the $\lambda\phi^4$ model \cite{Katz}.
\end{itemize}

\subsubsection*{Black holes and string theory}
Recent studies have shown connections between black holes and chaos by considering the OTOC \cite{BHChaos} and also in the S-matrix of particles interacting with a BH horizon \cite{Polchinski}. Chaotic behavior is also observed in the decay of highly excited strings \cite{Rosenhausstrings,Cobistrings}. It might be interesting to see connections between the two subjects, based on the BH-string correspondence \cite{HorowitzPolchinski,ChenMaldaWitten}.

\subsubsection*{New measures for mixed ensemble states}
We would like to explore more methods for determining whether a system is in a mixed state of e.g. integrable and GOE statistics. We suspect that the a linear combination of the pair correlations shown in section \ref{pair-correlations} for the different ensembles could indicate a mixed state with a single interpolating parameter.

\section{Acknowledgement}
 We would like to thank  A. Gaikwad  who took part in the early stages of this project and for  useful conversations. 
 J.S would like to thank M. Bianchi, A. Bhattacharya,  M. Firrotta, A. Jana  and D. Weissman for useful discussions. 
 This work  was supported in part by a grant 01034816 titled “String theory reloaded-
from fundamental questions to applications” of the “Planning and budgeting committee”.

\bibliographystyle{apsrev4-2}
\bibliography{Proposal}

\begin{thebibliography}{63}%
\makeatletter
\providecommand \@ifxundefined [1]{%
 \@ifx{#1\undefined}
}%
\providecommand \@ifnum [1]{%
 \ifnum #1\expandafter \@firstoftwo
 \else \expandafter \@secondoftwo
 \fi
}%
\providecommand \@ifx [1]{%
 \ifx #1\expandafter \@firstoftwo
 \else \expandafter \@secondoftwo
 \fi
}%
\providecommand \natexlab [1]{#1}%
\providecommand \enquote  [1]{``#1''}%
\providecommand \bibnamefont  [1]{#1}%
\providecommand \bibfnamefont [1]{#1}%
\providecommand \citenamefont [1]{#1}%
\providecommand \href@noop [0]{\@secondoftwo}%
\providecommand \href [0]{\begingroup \@sanitize@url \@href}%
\providecommand \@href[1]{\@@startlink{#1}\@@href}%
\providecommand \@@href[1]{\endgroup#1\@@endlink}%
\providecommand \@sanitize@url [0]{\catcode `\\12\catcode `\$12\catcode `\&12\catcode `\#12\catcode `\^12\catcode `\_12\catcode `\%12\relax}%
\providecommand \@@startlink[1]{}%
\providecommand \@@endlink[0]{}%
\providecommand \url  [0]{\begingroup\@sanitize@url \@url }%
\providecommand \@url [1]{\endgroup\@href {#1}{\urlprefix }}%
\providecommand \urlprefix  [0]{URL }%
\providecommand \Eprint [0]{\href }%
\providecommand \doibase [0]{https://doi.org/}%
\providecommand \selectlanguage [0]{\@gobble}%
\providecommand \bibinfo  [0]{\@secondoftwo}%
\providecommand \bibfield  [0]{\@secondoftwo}%
\providecommand \translation [1]{[#1]}%
\providecommand \BibitemOpen [0]{}%
\providecommand \bibitemStop [0]{}%
\providecommand \bibitemNoStop [0]{.\EOS\space}%
\providecommand \EOS [0]{\spacefactor3000\relax}%
\providecommand \BibitemShut  [1]{\csname bibitem#1\endcsname}%
\let\auto@bib@innerbib\@empty
\bibitem [{\citenamefont {Brandino}\ \emph {et~al.}(2010)\citenamefont {Brandino}, \citenamefont {Konik},\ and\ \citenamefont {Mussardo}}]{Brandino}%
  \BibitemOpen
  \bibfield  {author} {\bibinfo {author} {\bibfnamefont {G.~P.}\ \bibnamefont {Brandino}}, \bibinfo {author} {\bibfnamefont {R.~M.}\ \bibnamefont {Konik}},\ and\ \bibinfo {author} {\bibfnamefont {G.}~\bibnamefont {Mussardo}},\ }\href {https://doi.org/10.1088/1742-5468/2010/07/P07013} {\bibfield  {journal} {\bibinfo  {journal} {J. Stat. Mech.}\ }\textbf {\bibinfo {volume} {1007}},\ \bibinfo {pages} {P07013} (\bibinfo {year} {2010})},\ \Eprint {https://arxiv.org/abs/1004.4844} {arXiv:1004.4844 [cond-mat.stat-mech]} \BibitemShut {NoStop}%
\bibitem [{\citenamefont {Srdinsek}\ \emph {et~al.}(2021)\citenamefont {Srdinsek}, \citenamefont {Prosen},\ and\ \citenamefont {Sotiriadis}}]{Prosen}%
  \BibitemOpen
  \bibfield  {author} {\bibinfo {author} {\bibfnamefont {M.}~\bibnamefont {Srdinsek}}, \bibinfo {author} {\bibfnamefont {T.}~\bibnamefont {Prosen}},\ and\ \bibinfo {author} {\bibfnamefont {S.}~\bibnamefont {Sotiriadis}},\ }\href {https://doi.org/https://doi.org/10.48550/arXiv.2012.08505} {\bibfield  {journal} {\bibinfo  {journal} {Phys. Rev. Lett. 126,}\ } (\bibinfo {year} {2021})}\BibitemShut {NoStop}%
\bibitem [{\citenamefont {Delacretaz}\ \emph {et~al.}(2022)\citenamefont {Delacretaz}, \citenamefont {Fitzpatrick}, \citenamefont {Katz},\ and\ \citenamefont {Walters}}]{Katz}%
  \BibitemOpen
  \bibfield  {author} {\bibinfo {author} {\bibfnamefont {L.}~\bibnamefont {Delacretaz}}, \bibinfo {author} {\bibfnamefont {A.}~\bibnamefont {Fitzpatrick}}, \bibinfo {author} {\bibfnamefont {E.}~\bibnamefont {Katz}},\ and\ \bibinfo {author} {\bibfnamefont {M.}~\bibnamefont {Walters}},\ }\href {https://doi.org/https://doi.org/10.48550/arXiv.2207.11261} {\bibfield  {journal} {\bibinfo  {journal} {Phys.Rev.D57:2557-2563}\ } (\bibinfo {year} {2022})}\BibitemShut {NoStop}%
\bibitem [{\citenamefont {Cotler}\ \emph {et~al.}(2015)\citenamefont {Cotler}, \citenamefont {Gur-Ari}, \citenamefont {Hanada}, \citenamefont {Polchinski}, \citenamefont {Saad}, \citenamefont {Shenker}, \citenamefont {Stanford}, \citenamefont {Streicher},\ and\ \citenamefont {Tezuka}}]{BHChaos}%
  \BibitemOpen
  \bibfield  {author} {\bibinfo {author} {\bibfnamefont {J.}~\bibnamefont {Cotler}}, \bibinfo {author} {\bibfnamefont {G.}~\bibnamefont {Gur-Ari}}, \bibinfo {author} {\bibfnamefont {M.}~\bibnamefont {Hanada}}, \bibinfo {author} {\bibfnamefont {J.}~\bibnamefont {Polchinski}}, \bibinfo {author} {\bibfnamefont {P.}~\bibnamefont {Saad}}, \bibinfo {author} {\bibfnamefont {S.}~\bibnamefont {Shenker}}, \bibinfo {author} {\bibfnamefont {D.}~\bibnamefont {Stanford}}, \bibinfo {author} {\bibfnamefont {A.}~\bibnamefont {Streicher}},\ and\ \bibinfo {author} {\bibfnamefont {M.}~\bibnamefont {Tezuka}},\ }\href {https://doi.org/https://doi.org/10.48550/arXiv.1611.04650} {\bibfield  {journal} {\bibinfo  {journal} {JHEP 1705:118}\ } (\bibinfo {year} {2015})}\BibitemShut {NoStop}%
\bibitem [{\citenamefont {Rabinovici}\ \emph {et~al.}(2022)\citenamefont {Rabinovici}, \citenamefont {Sánchez-Garrido}, \citenamefont {Shir},\ and\ \citenamefont {Sonner}}]{Rabinovici}%
  \BibitemOpen
  \bibfield  {author} {\bibinfo {author} {\bibfnamefont {E.}~\bibnamefont {Rabinovici}}, \bibinfo {author} {\bibfnamefont {A.}~\bibnamefont {Sánchez-Garrido}}, \bibinfo {author} {\bibfnamefont {R.}~\bibnamefont {Shir}},\ and\ \bibinfo {author} {\bibfnamefont {J.}~\bibnamefont {Sonner}},\ }\href {https://doi.org/https://doi.org/10.48550/arXiv.2207.07701} {\bibfield  {journal} {\bibinfo  {journal} {Phys. Rev. Lett. 126,}\ } (\bibinfo {year} {2022})}\BibitemShut {NoStop}%
\bibitem [{\citenamefont {Avdoshkin}\ \emph {et~al.}(2024)\citenamefont {Avdoshkin}, \citenamefont {Dymarsky},\ and\ \citenamefont {Smolkin}}]{Avdoshkin}%
  \BibitemOpen
  \bibfield  {author} {\bibinfo {author} {\bibfnamefont {A.}~\bibnamefont {Avdoshkin}}, \bibinfo {author} {\bibfnamefont {A.}~\bibnamefont {Dymarsky}},\ and\ \bibinfo {author} {\bibfnamefont {M.}~\bibnamefont {Smolkin}},\ }\href {https://doi.org/10.1007/JHEP06(2024)066} {\bibfield  {journal} {\bibinfo  {journal} {JHEP}\ }\textbf {\bibinfo {volume} {06}},\ \bibinfo {pages} {066}},\ \Eprint {https://arxiv.org/abs/2212.14429} {arXiv:2212.14429 [hep-th]} \BibitemShut {NoStop}%
\bibitem [{\citenamefont {Balasubramanian}\ \emph {et~al.}(2025)\citenamefont {Balasubramanian}, \citenamefont {Das}, \citenamefont {Erdmenger},\ and\ \citenamefont {Xian}}]{Balasubramanian}%
  \BibitemOpen
  \bibfield  {author} {\bibinfo {author} {\bibfnamefont {V.}~\bibnamefont {Balasubramanian}}, \bibinfo {author} {\bibfnamefont {R.~N.}\ \bibnamefont {Das}}, \bibinfo {author} {\bibfnamefont {J.}~\bibnamefont {Erdmenger}},\ and\ \bibinfo {author} {\bibfnamefont {Z.-Y.}\ \bibnamefont {Xian}},\ }\href {https://doi.org/10.1088/1742-5468/adba41} {\bibfield  {journal} {\bibinfo  {journal} {J. Stat. Mech.}\ }\textbf {\bibinfo {volume} {2025}},\ \bibinfo {pages} {033202} (\bibinfo {year} {2025})},\ \Eprint {https://arxiv.org/abs/2407.11114} {arXiv:2407.11114 [hep-th]} \BibitemShut {NoStop}%
\bibitem [{\citenamefont {Gross}\ and\ \citenamefont {Rosenhaus}(2021)}]{Gross:2021gsj}%
  \BibitemOpen
  \bibfield  {author} {\bibinfo {author} {\bibfnamefont {D.~J.}\ \bibnamefont {Gross}}\ and\ \bibinfo {author} {\bibfnamefont {V.}~\bibnamefont {Rosenhaus}},\ }\href {https://doi.org/10.1007/JHEP05(2021)048} {\bibfield  {journal} {\bibinfo  {journal} {JHEP}\ }\textbf {\bibinfo {volume} {05}},\ \bibinfo {pages} {048}},\ \Eprint {https://arxiv.org/abs/2103.15301} {arXiv:2103.15301 [hep-th]} \BibitemShut {NoStop}%
\bibitem [{\citenamefont {Rosenhaus}(2021{\natexlab{a}})}]{Rosenhausstrings}%
  \BibitemOpen
  \bibfield  {author} {\bibinfo {author} {\bibfnamefont {V.}~\bibnamefont {Rosenhaus}},\ }\href {https://doi.org/https://doi.org/10.48550/arXiv.2112.10269} {\  (\bibinfo {year} {2021}{\natexlab{a}})}\BibitemShut {NoStop}%
\bibitem [{\citenamefont {Bianchi}\ \emph {et~al.}(2022)\citenamefont {Bianchi}, \citenamefont {Firrotta}, \citenamefont {Sonnenschein},\ and\ \citenamefont {Weissman}}]{Bianchi:2022mhs}%
  \BibitemOpen
  \bibfield  {author} {\bibinfo {author} {\bibfnamefont {M.}~\bibnamefont {Bianchi}}, \bibinfo {author} {\bibfnamefont {M.}~\bibnamefont {Firrotta}}, \bibinfo {author} {\bibfnamefont {J.}~\bibnamefont {Sonnenschein}},\ and\ \bibinfo {author} {\bibfnamefont {D.}~\bibnamefont {Weissman}},\ }\href {https://doi.org/10.1103/PhysRevLett.129.261601} {\bibfield  {journal} {\bibinfo  {journal} {Phys. Rev. Lett.}\ }\textbf {\bibinfo {volume} {129}},\ \bibinfo {pages} {261601} (\bibinfo {year} {2022})},\ \Eprint {https://arxiv.org/abs/2207.13112} {arXiv:2207.13112 [hep-th]} \BibitemShut {NoStop}%
\bibitem [{\citenamefont {Bianchi}\ \emph {et~al.}(2023)\citenamefont {Bianchi}, \citenamefont {Firrotta}, \citenamefont {Sonnenschein},\ and\ \citenamefont {Weissman}}]{Cobistrings}%
  \BibitemOpen
  \bibfield  {author} {\bibinfo {author} {\bibfnamefont {M.}~\bibnamefont {Bianchi}}, \bibinfo {author} {\bibfnamefont {M.}~\bibnamefont {Firrotta}}, \bibinfo {author} {\bibfnamefont {J.}~\bibnamefont {Sonnenschein}},\ and\ \bibinfo {author} {\bibfnamefont {D.}~\bibnamefont {Weissman}},\ }\href {https://doi.org/https://doi.org/10.48550/arXiv.2303.17233} {\  (\bibinfo {year} {2023})}\BibitemShut {NoStop}%
\bibitem [{\citenamefont {Bianchi}\ \emph {et~al.}(2024)\citenamefont {Bianchi}, \citenamefont {Firrotta}, \citenamefont {Sonnenschein},\ and\ \citenamefont {Weissman}}]{CobScFF}%
  \BibitemOpen
  \bibfield  {author} {\bibinfo {author} {\bibfnamefont {M.}~\bibnamefont {Bianchi}}, \bibinfo {author} {\bibfnamefont {M.}~\bibnamefont {Firrotta}}, \bibinfo {author} {\bibfnamefont {J.}~\bibnamefont {Sonnenschein}},\ and\ \bibinfo {author} {\bibfnamefont {D.}~\bibnamefont {Weissman}},\ }\href {https://doi.org/https://doi.org/10.48550/arXiv.2403.00713} {\  (\bibinfo {year} {2024})}\BibitemShut {NoStop}%
\bibitem [{\citenamefont {Negro}\ \emph {et~al.}(2023)\citenamefont {Negro}, \citenamefont {Popov},\ and\ \citenamefont {Sonnenschein}}]{Negro:2022hno}%
  \BibitemOpen
  \bibfield  {author} {\bibinfo {author} {\bibfnamefont {S.}~\bibnamefont {Negro}}, \bibinfo {author} {\bibfnamefont {F.~K.}\ \bibnamefont {Popov}},\ and\ \bibinfo {author} {\bibfnamefont {J.}~\bibnamefont {Sonnenschein}},\ }\href {https://doi.org/10.1103/PhysRevD.108.105024} {\bibfield  {journal} {\bibinfo  {journal} {Phys. Rev. D}\ }\textbf {\bibinfo {volume} {108}},\ \bibinfo {pages} {105024} (\bibinfo {year} {2023})},\ \Eprint {https://arxiv.org/abs/2211.14150} {arXiv:2211.14150 [hep-th]} \BibitemShut {NoStop}%
\bibitem [{\citenamefont {Dyson}(1962)}]{Dyson:1962b}%
  \BibitemOpen
  \bibfield  {author} {\bibinfo {author} {\bibfnamefont {F.~J.}\ \bibnamefont {Dyson}},\ }\href {https://doi.org/10.1063/1.1703774} {\bibfield  {journal} {\bibinfo  {journal} {Journal of Mathematical Physics}\ }\textbf {\bibinfo {volume} {3}},\ \bibinfo {pages} {157} (\bibinfo {year} {1962})},\ \Eprint {https://arxiv.org/abs/https://doi.org/10.1063/1.1703774} {https://doi.org/10.1063/1.1703774} \BibitemShut {NoStop}%
\bibitem [{\citenamefont {Bohigas}\ \emph {et~al.}(1984)\citenamefont {Bohigas}, \citenamefont {Giannoni},\ and\ \citenamefont {Schmit}}]{BGS}%
  \BibitemOpen
  \bibfield  {author} {\bibinfo {author} {\bibfnamefont {O.}~\bibnamefont {Bohigas}}, \bibinfo {author} {\bibfnamefont {M.}~\bibnamefont {Giannoni}},\ and\ \bibinfo {author} {\bibfnamefont {C.}~\bibnamefont {Schmit}},\ }\href {https://doi.org/https://doi.org/10.1103/PhysRevLett.52.1} {\bibfield  {journal} {\bibinfo  {journal} {Phys. Rev. Lett.}\ }\textbf {\bibinfo {volume} {52}} (\bibinfo {year} {1984})}\BibitemShut {NoStop}%
\bibitem [{\citenamefont {Berry}\ and\ \citenamefont {Tabor}(1977)}]{BerryTabor}%
  \BibitemOpen
  \bibfield  {author} {\bibinfo {author} {\bibfnamefont {M.}~\bibnamefont {Berry}}\ and\ \bibinfo {author} {\bibfnamefont {M.}~\bibnamefont {Tabor}},\ }\href {https://doi.org/https://doi.org/10.1098/rspa.1977.0140} {\bibfield  {journal} {\bibinfo  {journal} {Proc. R. Soc. Lond. A}\ }\textbf {\bibinfo {volume} {356}},\ \bibinfo {pages} {375–394} (\bibinfo {year} {1977})}\BibitemShut {NoStop}%
\bibitem [{\citenamefont {Brody}(1973)}]{Brody1}%
  \BibitemOpen
  \bibfield  {author} {\bibinfo {author} {\bibfnamefont {T.}~\bibnamefont {Brody}},\ }\href {https://doi.org/https://doi.org/10.1007/BF02727859} {\bibfield  {journal} {\bibinfo  {journal} {Lettere al Nuovo Cimento}\ }\textbf {\bibinfo {volume} {7}},\ \bibinfo {pages} {482–484} (\bibinfo {year} {1973})}\BibitemShut {NoStop}%
\bibitem [{\citenamefont {Brody}\ \emph {et~al.}(1981)\citenamefont {Brody}, \citenamefont {Flores}, \citenamefont {French}, \citenamefont {Mello}, \citenamefont {Pandey},\ and\ \citenamefont {Wong}}]{Brody2}%
  \BibitemOpen
  \bibfield  {author} {\bibinfo {author} {\bibfnamefont {T.}~\bibnamefont {Brody}}, \bibinfo {author} {\bibfnamefont {J.}~\bibnamefont {Flores}}, \bibinfo {author} {\bibfnamefont {J.}~\bibnamefont {French}}, \bibinfo {author} {\bibfnamefont {P.}~\bibnamefont {Mello}}, \bibinfo {author} {\bibfnamefont {A.}~\bibnamefont {Pandey}},\ and\ \bibinfo {author} {\bibfnamefont {S.}~\bibnamefont {Wong}},\ }\href {https://doi.org/https://doi.org/10.1103/RevModPhys.53.385} {\bibfield  {journal} {\bibinfo  {journal} {Rev. Mod. Phys.}\ }\textbf {\bibinfo {volume} {53}},\ \bibinfo {pages} {385} (\bibinfo {year} {1981})}\BibitemShut {NoStop}%
\bibitem [{\citenamefont {Izrailev}(1990)}]{Izrailev1}%
  \BibitemOpen
  \bibfield  {author} {\bibinfo {author} {\bibfnamefont {F.}~\bibnamefont {Izrailev}},\ }\href {https://doi.org/https://doi.org/10.1016/0370-1573(90)90067-C} {\bibfield  {journal} {\bibinfo  {journal} {Physics Reports}\ }\textbf {\bibinfo {volume} {196}},\ \bibinfo {pages} {299} (\bibinfo {year} {1990})}\BibitemShut {NoStop}%
\bibitem [{\citenamefont {Chirikov}\ \emph {et~al.}(1988)\citenamefont {Chirikov}, \citenamefont {Izrailev},\ and\ \citenamefont {Shepelyansky}}]{Izrailev2}%
  \BibitemOpen
  \bibfield  {author} {\bibinfo {author} {\bibfnamefont {B.}~\bibnamefont {Chirikov}}, \bibinfo {author} {\bibfnamefont {F.}~\bibnamefont {Izrailev}},\ and\ \bibinfo {author} {\bibfnamefont {D.}~\bibnamefont {Shepelyansky}},\ }\href {https://doi.org/https://doi.org/10.1016/S0167-2789(98)90011-2} {\bibfield  {journal} {\bibinfo  {journal} {Physica D: Nonlinear Phenomena}\ }\textbf {\bibinfo {volume} {33}},\ \bibinfo {pages} {77} (\bibinfo {year} {1988})}\BibitemShut {NoStop}%
\bibitem [{\citenamefont {Oganesyan}\ and\ \citenamefont {Huse}(2007)}]{OganesyanHuse}%
  \BibitemOpen
  \bibfield  {author} {\bibinfo {author} {\bibfnamefont {V.}~\bibnamefont {Oganesyan}}\ and\ \bibinfo {author} {\bibfnamefont {D.}~\bibnamefont {Huse}},\ }\href {https://doi.org/https://doi.org/10.1103/PhysRevB.75.155111} {\bibfield  {journal} {\bibinfo  {journal} {Phys. Rev. B}\ }\textbf {\bibinfo {volume} {75}} (\bibinfo {year} {2007})}\BibitemShut {NoStop}%
\bibitem [{\citenamefont {Atas}\ \emph {et~al.}(2013{\natexlab{a}})\citenamefont {Atas}, \citenamefont {Bogomolny}, \citenamefont {Giraud},\ and\ \citenamefont {Roux}}]{ABGR}%
  \BibitemOpen
  \bibfield  {author} {\bibinfo {author} {\bibfnamefont {Y.~Y.}\ \bibnamefont {Atas}}, \bibinfo {author} {\bibfnamefont {E.}~\bibnamefont {Bogomolny}}, \bibinfo {author} {\bibfnamefont {O.}~\bibnamefont {Giraud}},\ and\ \bibinfo {author} {\bibfnamefont {G.}~\bibnamefont {Roux}},\ }\href {https://doi.org/https://doi.org/10.1103/PhysRevLett.110.084101} {\bibfield  {journal} {\bibinfo  {journal} {Phys. Rev. Lett.}\ }\textbf {\bibinfo {volume} {110}} (\bibinfo {year} {2013}{\natexlab{a}})}\BibitemShut {NoStop}%
\bibitem [{\citenamefont {Atas}\ \emph {et~al.}(2013{\natexlab{b}})\citenamefont {Atas}, \citenamefont {Bogomolny}, \citenamefont {Giraud}, \citenamefont {Vivo},\ and\ \citenamefont {Vivo}}]{ABGVV}%
  \BibitemOpen
  \bibfield  {author} {\bibinfo {author} {\bibfnamefont {Y.~Y.}\ \bibnamefont {Atas}}, \bibinfo {author} {\bibfnamefont {E.}~\bibnamefont {Bogomolny}}, \bibinfo {author} {\bibfnamefont {O.}~\bibnamefont {Giraud}}, \bibinfo {author} {\bibfnamefont {E.}~\bibnamefont {Vivo}},\ and\ \bibinfo {author} {\bibfnamefont {P.}~\bibnamefont {Vivo}},\ }\href {https://doi.org/http://dx.doi.org/10.1002/andp.19053221004} {\bibfield  {journal} {\bibinfo  {journal} {J. Phys. A: Math. Theor.}\ }\textbf {\bibinfo {volume} {46}} (\bibinfo {year} {2013}{\natexlab{b}})}\BibitemShut {NoStop}%
\bibitem [{\citenamefont {Chenu}\ \emph {et~al.}(2023)\citenamefont {Chenu}, \citenamefont {Martinez-Azcona},\ and\ \citenamefont {Shir}}]{ChenuMartinezShir}%
  \BibitemOpen
  \bibfield  {author} {\bibinfo {author} {\bibfnamefont {A.}~\bibnamefont {Chenu}}, \bibinfo {author} {\bibfnamefont {P.}~\bibnamefont {Martinez-Azcona}},\ and\ \bibinfo {author} {\bibfnamefont {R.}~\bibnamefont {Shir}},\ }\href {https://doi.org/https://doi.org/10.48550/arXiv.2311.09292} {\  (\bibinfo {year} {2023})}\BibitemShut {NoStop}%
\bibitem [{\citenamefont {Fox}\ and\ \citenamefont {Kahn}(1964)}]{FoxKhan}%
  \BibitemOpen
  \bibfield  {author} {\bibinfo {author} {\bibfnamefont {D.}~\bibnamefont {Fox}}\ and\ \bibinfo {author} {\bibfnamefont {P.}~\bibnamefont {Kahn}},\ }\href {https://doi.org/https://doi.org/10.1103/PhysRev.134.B1151} {\bibfield  {journal} {\bibinfo  {journal} {Phys. Rev.}\ }\textbf {\bibinfo {volume} {134}} (\bibinfo {year} {1964})}\BibitemShut {NoStop}%
\bibitem [{\citenamefont {Engel}\ \emph {et~al.}(1998)\citenamefont {Engel}, \citenamefont {Main},\ and\ \citenamefont {Wunner}}]{EngelMainWunner}%
  \BibitemOpen
  \bibfield  {author} {\bibinfo {author} {\bibfnamefont {D.}~\bibnamefont {Engel}}, \bibinfo {author} {\bibfnamefont {J.}~\bibnamefont {Main}},\ and\ \bibinfo {author} {\bibfnamefont {G.}~\bibnamefont {Wunner}},\ }\href {https://doi.org/10.1088/0305-4470/31/33/007} {\bibfield  {journal} {\bibinfo  {journal} {J. Phys. A: Math. Gen.}\ }\textbf {\bibinfo {volume} {31}} (\bibinfo {year} {1998})}\BibitemShut {NoStop}%
\bibitem [{\citenamefont {Abul-Magd}\ and\ \citenamefont {Simbel}(1999)}]{AbulMagdSimbel}%
  \BibitemOpen
  \bibfield  {author} {\bibinfo {author} {\bibfnamefont {A.}~\bibnamefont {Abul-Magd}}\ and\ \bibinfo {author} {\bibfnamefont {M.}~\bibnamefont {Simbel}},\ }\href {https://doi.org/https://doi.org/10.1103/PhysRevE.60.5371} {\bibfield  {journal} {\bibinfo  {journal} {Phys. Rev. E}\ }\textbf {\bibinfo {volume} {60}} (\bibinfo {year} {1999})}\BibitemShut {NoStop}%
\bibitem [{\citenamefont {Rao}(2020)}]{Rao}%
  \BibitemOpen
  \bibfield  {author} {\bibinfo {author} {\bibfnamefont {W.}~\bibnamefont {Rao}},\ }\href {https://doi.org/https://doi.org/10.1103/PhysRevB.102.054202} {\bibfield  {journal} {\bibinfo  {journal} {Phys. Rev. B}\ }\textbf {\bibinfo {volume} {102}} (\bibinfo {year} {2020})}\BibitemShut {NoStop}%
\bibitem [{\citenamefont {Tekur}\ \emph {et~al.}(2018)\citenamefont {Tekur}, \citenamefont {Bhosale},\ and\ \citenamefont {Santhanam}}]{HarishniTekur}%
  \BibitemOpen
  \bibfield  {author} {\bibinfo {author} {\bibfnamefont {S.~H.}\ \bibnamefont {Tekur}}, \bibinfo {author} {\bibfnamefont {U.}~\bibnamefont {Bhosale}},\ and\ \bibinfo {author} {\bibfnamefont {M.}~\bibnamefont {Santhanam}},\ }\href {https://doi.org/https://doi.org/10.1103/PhysRevB.98.104305} {\bibfield  {journal} {\bibinfo  {journal} {Phys. Rev. B}\ }\textbf {\bibinfo {volume} {98}} (\bibinfo {year} {2018})}\BibitemShut {NoStop}%
\bibitem [{\citenamefont {Guhr}\ \emph {et~al.}(1998)\citenamefont {Guhr}, \citenamefont {Müller–Groeling},\ and\ \citenamefont {Weidenmüller}}]{Weidenmuller}%
  \BibitemOpen
  \bibfield  {author} {\bibinfo {author} {\bibfnamefont {T.}~\bibnamefont {Guhr}}, \bibinfo {author} {\bibfnamefont {A.}~\bibnamefont {Müller–Groeling}},\ and\ \bibinfo {author} {\bibfnamefont {H.}~\bibnamefont {Weidenmüller}},\ }\href {https://doi.org/https://doi.org/10.1016/S0370-1573(97)00088-4} {\bibfield  {journal} {\bibinfo  {journal} {Physics Reports}\ }\textbf {\bibinfo {volume} {299}},\ \bibinfo {pages} {189} (\bibinfo {year} {1998})}\BibitemShut {NoStop}%
\bibitem [{\citenamefont {Montgomery}(1973)}]{Montgomery}%
  \BibitemOpen
  \bibfield  {author} {\bibinfo {author} {\bibfnamefont {H.}~\bibnamefont {Montgomery}},\ }\href {https://doi.org/https://doi.org/10.1016/S0370-1573(97)00088-4} {\bibfield  {journal} {\bibinfo  {journal} {Proc. Sympos. Pure Math.}\ }\textbf {\bibinfo {volume} {24}} (\bibinfo {year} {1973})}\BibitemShut {NoStop}%
\bibitem [{\citenamefont {Odlyzko}(1987)}]{Odlyzko}%
  \BibitemOpen
  \bibfield  {author} {\bibinfo {author} {\bibfnamefont {A.~M.}\ \bibnamefont {Odlyzko}},\ }\href {http://www.jstor.org/stable/2007890} {\bibfield  {journal} {\bibinfo  {journal} {Mathematics of Computation}\ }\textbf {\bibinfo {volume} {48}},\ \bibinfo {pages} {273} (\bibinfo {year} {1987})}\BibitemShut {NoStop}%
\bibitem [{\citenamefont {Stöckmann}\ and\ \citenamefont {Gutzwiller}(2000)}]{stockmann}%
  \BibitemOpen
  \bibfield  {author} {\bibinfo {author} {\bibfnamefont {H.-J.}\ \bibnamefont {Stöckmann}}\ and\ \bibinfo {author} {\bibfnamefont {M.}~\bibnamefont {Gutzwiller}},\ }\href {https://doi.org/10.1119/1.19544} {\bibfield  {journal} {\bibinfo  {journal} {American Journal of Physics}\ }\textbf {\bibinfo {volume} {68}},\ \bibinfo {pages} {777} (\bibinfo {year} {2000})}\BibitemShut {NoStop}%
\bibitem [{\citenamefont {D'Alessio}\ \emph {et~al.}(2016)\citenamefont {D'Alessio}, \citenamefont {Kafri}, \citenamefont {Polkovnikov},\ and\ \citenamefont {Rigol}}]{Kafri}%
  \BibitemOpen
  \bibfield  {author} {\bibinfo {author} {\bibfnamefont {L.}~\bibnamefont {D'Alessio}}, \bibinfo {author} {\bibfnamefont {Y.}~\bibnamefont {Kafri}}, \bibinfo {author} {\bibfnamefont {A.}~\bibnamefont {Polkovnikov}},\ and\ \bibinfo {author} {\bibfnamefont {M.}~\bibnamefont {Rigol}},\ }\href {https://doi.org/https://doi.org/10.48550/arXiv.1509.06411} {\bibfield  {journal} {\bibinfo  {journal} {Adv. Phys.}\ }\textbf {\bibinfo {volume} {65}} (\bibinfo {year} {2016})}\BibitemShut {NoStop}%
\bibitem [{\citenamefont {Haake}\ \emph {et~al.}(2018)\citenamefont {Haake}, \citenamefont {Gnutzmann},\ and\ \citenamefont {Kuś}}]{Haake}%
  \BibitemOpen
  \bibfield  {author} {\bibinfo {author} {\bibfnamefont {F.}~\bibnamefont {Haake}}, \bibinfo {author} {\bibfnamefont {S.}~\bibnamefont {Gnutzmann}},\ and\ \bibinfo {author} {\bibfnamefont {M.}~\bibnamefont {Kuś}},\ }\href@noop {} {\emph {\bibinfo {title} {Quantum Signatures of Chaos}}}\ (\bibinfo  {publisher} {Springer Cham},\ \bibinfo {year} {2018})\BibitemShut {NoStop}%
\bibitem [{\citenamefont {Das}\ \emph {et~al.}(2024)\citenamefont {Das}, \citenamefont {Garg}, \citenamefont {Krishnan},\ and\ \citenamefont {Kundu}}]{SFFramp}%
  \BibitemOpen
  \bibfield  {author} {\bibinfo {author} {\bibfnamefont {S.}~\bibnamefont {Das}}, \bibinfo {author} {\bibfnamefont {S.}~\bibnamefont {Garg}}, \bibinfo {author} {\bibfnamefont {C.}~\bibnamefont {Krishnan}},\ and\ \bibinfo {author} {\bibfnamefont {A.}~\bibnamefont {Kundu}},\ }\href {https://doi.org/https://doi.org/10.48550/arXiv.2308.11704} {\  (\bibinfo {year} {2024})}\BibitemShut {NoStop}%
\bibitem [{\citenamefont {Haake}\ \emph {et~al.}(1967)\citenamefont {Haake}, \citenamefont {Gnutzmann},\ and\ \citenamefont {Kuś}}]{Mehta}%
  \BibitemOpen
  \bibfield  {author} {\bibinfo {author} {\bibfnamefont {F.}~\bibnamefont {Haake}}, \bibinfo {author} {\bibfnamefont {S.}~\bibnamefont {Gnutzmann}},\ and\ \bibinfo {author} {\bibfnamefont {M.}~\bibnamefont {Kuś}},\ }\href@noop {} {\emph {\bibinfo {title} {Random Matrices and the Statistical Theory of Energy Levels}}}\ (\bibinfo  {publisher} {Academic Press},\ \bibinfo {year} {1967})\BibitemShut {NoStop}%
\bibitem [{\citenamefont {Persson}\ and\ \citenamefont {Aberg}(1995)}]{PhysRevE:52:148}%
  \BibitemOpen
  \bibfield  {author} {\bibinfo {author} {\bibfnamefont {P.}~\bibnamefont {Persson}}\ and\ \bibinfo {author} {\bibfnamefont {S.}~\bibnamefont {Aberg}},\ }\href {https://doi.org/10.1103/PhysRevE.52.148} {\bibfield  {journal} {\bibinfo  {journal} {Phys. Rev. E}\ }\textbf {\bibinfo {volume} {52}},\ \bibinfo {pages} {148} (\bibinfo {year} {1995})}\BibitemShut {NoStop}%
\bibitem [{\citenamefont {Serbyn}\ and\ \citenamefont {Moore}(2016)}]{PhysRevB.93.041424}%
  \BibitemOpen
  \bibfield  {author} {\bibinfo {author} {\bibfnamefont {M.}~\bibnamefont {Serbyn}}\ and\ \bibinfo {author} {\bibfnamefont {J.~E.}\ \bibnamefont {Moore}},\ }\href {https://doi.org/10.1103/PhysRevB.93.041424} {\bibfield  {journal} {\bibinfo  {journal} {Phys. Rev. B}\ }\textbf {\bibinfo {volume} {93}},\ \bibinfo {pages} {041424} (\bibinfo {year} {2016})}\BibitemShut {NoStop}%
\bibitem [{\citenamefont {Sierant}\ \emph {et~al.}(2020)\citenamefont {Sierant}, \citenamefont {Delande},\ and\ \citenamefont {Zakrzewski}}]{Sierant_2020}%
  \BibitemOpen
  \bibfield  {author} {\bibinfo {author} {\bibfnamefont {P.}~\bibnamefont {Sierant}}, \bibinfo {author} {\bibfnamefont {D.}~\bibnamefont {Delande}},\ and\ \bibinfo {author} {\bibfnamefont {J.}~\bibnamefont {Zakrzewski}},\ }\bibfield  {journal} {\bibinfo  {journal} {Physical Review Letters}\ }\textbf {\bibinfo {volume} {124}},\ \href {https://doi.org/10.1103/physrevlett.124.186601} {10.1103/physrevlett.124.186601} (\bibinfo {year} {2020})\BibitemShut {NoStop}%
\bibitem [{\citenamefont {Bhattacharya}\ \emph {et~al.}(2025)\citenamefont {Bhattacharya}, \citenamefont {Jana},\ and\ \citenamefont {J.Sonnenschein}}]{Bhath:25}%
  \BibitemOpen
  \bibfield  {author} {\bibinfo {author} {\bibfnamefont {A.}~\bibnamefont {Bhattacharya}}, \bibinfo {author} {\bibfnamefont {A.}~\bibnamefont {Jana}},\ and\ \bibinfo {author} {\bibnamefont {J.Sonnenschein}},\ }\href@noop {} {\  (\bibinfo {year} {2025})}\BibitemShut {NoStop}%
\bibitem [{\citenamefont {Rosenhaus}(2021{\natexlab{b}})}]{Rosenhaus}%
  \BibitemOpen
  \bibfield  {author} {\bibinfo {author} {\bibfnamefont {V.}~\bibnamefont {Rosenhaus}},\ }\href {https://doi.org/https://doi.org/10.48550/arXiv.2003.07381} {\bibfield  {journal} {\bibinfo  {journal} {Phys. Rev. Lett. 127, 021601}\ } (\bibinfo {year} {2021}{\natexlab{b}})}\BibitemShut {NoStop}%
\bibitem [{\citenamefont {Yurov}\ and\ \citenamefont {Zamolodchikov}(1989)}]{TCSA}%
  \BibitemOpen
  \bibfield  {author} {\bibinfo {author} {\bibfnamefont {V.}~\bibnamefont {Yurov}}\ and\ \bibinfo {author} {\bibfnamefont {A.}~\bibnamefont {Zamolodchikov}},\ }\href {https://doi.org/https://doi.org/10.1142/S0217751X9000218X} {\bibfield  {journal} {\bibinfo  {journal} {International Journal of Modern Physics A}\ }\textbf {\bibinfo {volume} {5}} (\bibinfo {year} {1989})}\BibitemShut {NoStop}%
\bibitem [{\citenamefont {Anand}\ \emph {et~al.}(2020)\citenamefont {Anand}, \citenamefont {Fitzpatrick}, \citenamefont {Katz}, \citenamefont {Khandker}, \citenamefont {Walters},\ and\ \citenamefont {Xin}}]{LightconeTrunc}%
  \BibitemOpen
  \bibfield  {author} {\bibinfo {author} {\bibfnamefont {N.}~\bibnamefont {Anand}}, \bibinfo {author} {\bibfnamefont {A.}~\bibnamefont {Fitzpatrick}}, \bibinfo {author} {\bibfnamefont {E.}~\bibnamefont {Katz}}, \bibinfo {author} {\bibfnamefont {Z.}~\bibnamefont {Khandker}}, \bibinfo {author} {\bibfnamefont {M.}~\bibnamefont {Walters}},\ and\ \bibinfo {author} {\bibfnamefont {Y.}~\bibnamefont {Xin}},\ }\href {https://doi.org/https://doi.org/10.48550/arXiv.2005.13544} {\  (\bibinfo {year} {2020})}\BibitemShut {NoStop}%
\bibitem [{\citenamefont {Belin}\ \emph {et~al.}(2024)\citenamefont {Belin}, \citenamefont {de~Boer}, \citenamefont {Jafferis}, \citenamefont {Nayak},\ and\ \citenamefont {Sonner}}]{Sonner}%
  \BibitemOpen
  \bibfield  {author} {\bibinfo {author} {\bibfnamefont {A.}~\bibnamefont {Belin}}, \bibinfo {author} {\bibfnamefont {J.}~\bibnamefont {de~Boer}}, \bibinfo {author} {\bibfnamefont {D.}~\bibnamefont {Jafferis}}, \bibinfo {author} {\bibfnamefont {P.}~\bibnamefont {Nayak}},\ and\ \bibinfo {author} {\bibfnamefont {J.}~\bibnamefont {Sonner}},\ }\href {https://doi.org/https://doi.org/10.48550/arXiv.2308.03829} {\bibfield  {journal} {\bibinfo  {journal} {Phys.Rev.D57:2557-2563}\ } (\bibinfo {year} {2024})}\BibitemShut {NoStop}%
\bibitem [{\citenamefont {Goddard}\ and\ \citenamefont {Olive}(1986)}]{WZW}%
  \BibitemOpen
  \bibfield  {author} {\bibinfo {author} {\bibfnamefont {P.}~\bibnamefont {Goddard}}\ and\ \bibinfo {author} {\bibfnamefont {D.}~\bibnamefont {Olive}},\ }\href {https://doi.org/https://doi.org/10.1142/S0217751X86000149} {\bibfield  {journal} {\bibinfo  {journal} {International Journal of Modern Physics AVol. 01, No. 02}\ } (\bibinfo {year} {1986})}\BibitemShut {NoStop}%
\bibitem [{\citenamefont {'t~Hooft}(1974)}]{tHooft}%
  \BibitemOpen
  \bibfield  {author} {\bibinfo {author} {\bibfnamefont {G.}~\bibnamefont {'t~Hooft}},\ }\href {https://doi.org/10.1016/0550-3213(74)90088-1} {\bibfield  {journal} {\bibinfo  {journal} {Nucl.Phys.B 75}\ } (\bibinfo {year} {1974})}\BibitemShut {NoStop}%
\bibitem [{\citenamefont {Frishman}\ and\ \citenamefont {Sonnenschein}(1993)}]{Bosonization}%
  \BibitemOpen
  \bibfield  {author} {\bibinfo {author} {\bibfnamefont {Y.}~\bibnamefont {Frishman}}\ and\ \bibinfo {author} {\bibfnamefont {J.}~\bibnamefont {Sonnenschein}},\ }\href {https://doi.org/https://doi.org/10.48550/arXiv.hep-th/9207017} {\bibfield  {journal} {\bibinfo  {journal} {Phys.Rept.223:309-348}\ } (\bibinfo {year} {1993})}\BibitemShut {NoStop}%
\bibitem [{\citenamefont {Kovacs}(1999)}]{SYM}%
  \BibitemOpen
  \bibfield  {author} {\bibinfo {author} {\bibfnamefont {S.}~\bibnamefont {Kovacs}},\ }\href {https://doi.org/arXiv:hep-th/9908171} {\  (\bibinfo {year} {1999})}\BibitemShut {NoStop}%
\bibitem [{\citenamefont {Polchinski}(2015)}]{Polchinski}%
  \BibitemOpen
  \bibfield  {author} {\bibinfo {author} {\bibfnamefont {J.}~\bibnamefont {Polchinski}},\ }\href {https://doi.org/https://doi.org/10.48550/arXiv.1505.08108} {\bibfield  {journal} {\bibinfo  {journal} {Phys.Rept.223:309-348}\ } (\bibinfo {year} {2015})}\BibitemShut {NoStop}%
\bibitem [{\citenamefont {Horowitz}\ and\ \citenamefont {Polchinski}(1998)}]{HorowitzPolchinski}%
  \BibitemOpen
  \bibfield  {author} {\bibinfo {author} {\bibfnamefont {G.}~\bibnamefont {Horowitz}}\ and\ \bibinfo {author} {\bibfnamefont {J.}~\bibnamefont {Polchinski}},\ }\href {https://doi.org/https://doi.org/10.48550/arXiv.hep-th/9707170} {\bibfield  {journal} {\bibinfo  {journal} {Phys.Rev.D57:2557-2563}\ } (\bibinfo {year} {1998})}\BibitemShut {NoStop}%
\bibitem [{\citenamefont {Chen}\ \emph {et~al.}(2021)\citenamefont {Chen}, \citenamefont {Maldacena},\ and\ \citenamefont {Witten}}]{ChenMaldaWitten}%
  \BibitemOpen
  \bibfield  {author} {\bibinfo {author} {\bibfnamefont {Y.}~\bibnamefont {Chen}}, \bibinfo {author} {\bibfnamefont {J.}~\bibnamefont {Maldacena}},\ and\ \bibinfo {author} {\bibfnamefont {E.}~\bibnamefont {Witten}},\ }\href {https://doi.org/https://doi.org/10.48550/arXiv.2109.08563} {\bibfield  {journal} {\bibinfo  {journal} {Phys.Rev.D57:2557-2563}\ } (\bibinfo {year} {2021})}\BibitemShut {NoStop}%
\bibitem [{\citenamefont {Yurov}\ and\ \citenamefont {Zamolodchikov}(1990)}]{Yurov:1989yu}%
  \BibitemOpen
  \bibfield  {author} {\bibinfo {author} {\bibfnamefont {V.~P.}\ \bibnamefont {Yurov}}\ and\ \bibinfo {author} {\bibfnamefont {A.~B.}\ \bibnamefont {Zamolodchikov}},\ }\href {https://doi.org/10.1142/S0217751X9000218X} {\bibfield  {journal} {\bibinfo  {journal} {Int. J. Mod. Phys. A}\ }\textbf {\bibinfo {volume} {5}},\ \bibinfo {pages} {3221} (\bibinfo {year} {1990})}\BibitemShut {NoStop}%
\bibitem [{\citenamefont {Hogervorst}\ \emph {et~al.}(2015)\citenamefont {Hogervorst}, \citenamefont {Rychkov},\ and\ \citenamefont {van Rees}}]{Hogervorst_2015}%
  \BibitemOpen
  \bibfield  {author} {\bibinfo {author} {\bibfnamefont {M.}~\bibnamefont {Hogervorst}}, \bibinfo {author} {\bibfnamefont {S.}~\bibnamefont {Rychkov}},\ and\ \bibinfo {author} {\bibfnamefont {B.~C.}\ \bibnamefont {van Rees}},\ }\bibfield  {journal} {\bibinfo  {journal} {Physical Review D}\ }\textbf {\bibinfo {volume} {91}},\ \href {https://doi.org/10.1103/physrevd.91.025005} {10.1103/physrevd.91.025005} (\bibinfo {year} {2015})\BibitemShut {NoStop}%
\bibitem [{\citenamefont {Konik}\ and\ \citenamefont {Adamov}(2007)}]{Konik_2007}%
  \BibitemOpen
  \bibfield  {author} {\bibinfo {author} {\bibfnamefont {R.~M.}\ \bibnamefont {Konik}}\ and\ \bibinfo {author} {\bibfnamefont {Y.}~\bibnamefont {Adamov}},\ }\bibfield  {journal} {\bibinfo  {journal} {Physical Review Letters}\ }\textbf {\bibinfo {volume} {98}},\ \href {https://doi.org/10.1103/physrevlett.98.147205} {10.1103/physrevlett.98.147205} (\bibinfo {year} {2007})\BibitemShut {NoStop}%
\bibitem [{\citenamefont {Cataliotti}\ \emph {et~al.}(2001)\citenamefont {Cataliotti}, \citenamefont {Burger}, \citenamefont {Fort}, \citenamefont {Maddaloni}, \citenamefont {Minardi}, \citenamefont {Trombettoni}, \citenamefont {Smerzi},\ and\ \citenamefont {Inguscio}}]{Josephson}%
  \BibitemOpen
  \bibfield  {author} {\bibinfo {author} {\bibfnamefont {F.~S.}\ \bibnamefont {Cataliotti}}, \bibinfo {author} {\bibfnamefont {S.}~\bibnamefont {Burger}}, \bibinfo {author} {\bibfnamefont {C.}~\bibnamefont {Fort}}, \bibinfo {author} {\bibfnamefont {P.}~\bibnamefont {Maddaloni}}, \bibinfo {author} {\bibfnamefont {F.}~\bibnamefont {Minardi}}, \bibinfo {author} {\bibfnamefont {A.}~\bibnamefont {Trombettoni}}, \bibinfo {author} {\bibfnamefont {A.}~\bibnamefont {Smerzi}},\ and\ \bibinfo {author} {\bibfnamefont {M.}~\bibnamefont {Inguscio}},\ }\href {https://doi.org/https://doi.org/10.48550/arXiv.cond-mat/0108117} {\bibfield  {journal} {\bibinfo  {journal} {Science, vol. 293}\ } (\bibinfo {year} {2001})}\BibitemShut {NoStop}%
\bibitem [{\citenamefont {Baxter}(1982)}]{Baxter}%
  \BibitemOpen
  \bibfield  {author} {\bibinfo {author} {\bibfnamefont {R.}~\bibnamefont {Baxter}},\ }\href@noop {} {\emph {\bibinfo {title} {Exactly Solved Models in Statistical Mechanics}}}\ (\bibinfo  {publisher} {Academic Press},\ \bibinfo {year} {1982})\BibitemShut {NoStop}%
\bibitem [{\citenamefont {Coleman}(1975)}]{Coleman}%
  \BibitemOpen
  \bibfield  {author} {\bibinfo {author} {\bibfnamefont {S.}~\bibnamefont {Coleman}},\ }\href {https://doi.org/https://doi.org/10.1103/PhysRevD.11.2088} {\bibfield  {journal} {\bibinfo  {journal} {Phys. Rev. D 11, 2088}\ } (\bibinfo {year} {1975})}\BibitemShut {NoStop}%
\bibitem [{\citenamefont {Delfino}\ and\ \citenamefont {Mussardo}(1998)}]{DelfinoMussardo}%
  \BibitemOpen
  \bibfield  {author} {\bibinfo {author} {\bibfnamefont {G.}~\bibnamefont {Delfino}}\ and\ \bibinfo {author} {\bibfnamefont {G.}~\bibnamefont {Mussardo}},\ }\href {https://doi.org/https://doi.org/10.48550/arXiv.hep-th/9709028} {\bibfield  {journal} {\bibinfo  {journal} {Nucl.Phys. B516}\ } (\bibinfo {year} {1998})}\BibitemShut {NoStop}%
\bibitem [{\citenamefont {Bajnok}\ \emph {et~al.}(2001)\citenamefont {Bajnok}, \citenamefont {Palla}, \citenamefont {Takacs},\ and\ \citenamefont {Wagner}}]{Banjok}%
  \BibitemOpen
  \bibfield  {author} {\bibinfo {author} {\bibfnamefont {Z.}~\bibnamefont {Bajnok}}, \bibinfo {author} {\bibfnamefont {L.}~\bibnamefont {Palla}}, \bibinfo {author} {\bibfnamefont {G.}~\bibnamefont {Takacs}},\ and\ \bibinfo {author} {\bibfnamefont {F.}~\bibnamefont {Wagner}},\ }\href {https://doi.org/https://doi.org/10.48550/arXiv.hep-th/0008066} {\bibfield  {journal} {\bibinfo  {journal} {Nucl.Phys. B601}\ } (\bibinfo {year} {2001})}\BibitemShut {NoStop}%
\bibitem [{\citenamefont {Erdos}\ and\ \citenamefont {Yau}(2011)}]{ErdosYau}%
  \BibitemOpen
  \bibfield  {author} {\bibinfo {author} {\bibfnamefont {L.}~\bibnamefont {Erdos}}\ and\ \bibinfo {author} {\bibfnamefont {H.}~\bibnamefont {Yau}},\ }\href {https://doi.org/https://doi.org/10.48550/arXiv.1106.4986} {\  (\bibinfo {year} {2011})}\BibitemShut {NoStop}%
\bibitem [{\citenamefont {Reichl}(2004)}]{Reichl}%
  \BibitemOpen
  \bibfield  {author} {\bibinfo {author} {\bibfnamefont {L.}~\bibnamefont {Reichl}},\ }\href@noop {} {\emph {\bibinfo {title} {The Transition to Chaos: Conservative Classical Systems and Quantum Manifestations}}}\ (\bibinfo  {publisher} {Springer New York, NY},\ \bibinfo {year} {2004})\BibitemShut {NoStop}%
\bibitem [{\citenamefont {Dumitriu}\ and\ \citenamefont {Edelman}(2002)}]{Dumitriu}%
  \BibitemOpen
  \bibfield  {author} {\bibinfo {author} {\bibfnamefont {I.}~\bibnamefont {Dumitriu}}\ and\ \bibinfo {author} {\bibfnamefont {A.}~\bibnamefont {Edelman}},\ }\href {https://doi.org/https://doi.org/10.48550/arXiv.math-ph/0206043} {\  (\bibinfo {year} {2002})}\BibitemShut {NoStop}%
\end{thebibliography}%

\appendix

\section{Truncated Conformal Space Approach} \label{TCSA}

Aside from calculations in RMT, the data used in this paper was taken from \cite{Prosen}, where the authors used the Truncated Conformal Space Approach (TCSA) to obtain the spectrum of Sine-Gordon and Double Sine-Gordon.
TCSA is a non-perturbative numerical method that can be used to study QFT, which takes advantage of the properties of Conformal Field Theory (CFT) by using its eigenstates as the basis of the calculations. It is part of the larger class of Hamiltonian truncations, and was first suggested by Yurov and Zamolodchikov in 1989 \cite{Yurov:1989yu} (for a more detailed account of TCSA, see for example \cite{Hogervorst_2015}). 
The basic idea is to start with a known CFT Hamiltonian and to perturb it with a relevant operator 

\begin{equation}
\label{eq:TCSA general}
H=H_{CFT}+g\int dx \mathcal{V}(x)
\end{equation}

Where $g$ is the coupling constant and $\mathcal{V}(x)$ is a local relevant operator.
This is done by choosing the eigenstates of the CFT as the basis $\ket{i}$ with the highest eigenvalue less than a fixed energy cutoff $E_{cut}$, and computing the matrix elements of the perturbations

\begin{equation}
\label{eq:TCSA matrix elements}
V_{ij}=\bra{i}\int dx \mathcal{V}(x)\ket{j}
\end{equation}

The result is then diagonalized numerically to obtain the new eigenvalues.\par
Though pretty conceptually straightforward, this method might be computationally pricy since the number of states tends to grow exponentially in the cutoff energy. This problem is tackled by exploiting the symmetries of the system so that the truncated Hamiltonian becomes block diagonal, and can be made even more efficient using RG flow methods (e.g. numerical RG \cite{Konik_2007}).\par
The validity of the method relies on the assumption that the cutoff energy is large enough so that the IR regime is unaffected by the cutoff scale, i.e.

\begin{equation}
\label{eq:TCSA convergence scale}
E_{cut}\gg|g|^{\frac{1}{2-\Delta}}
\end{equation}

where $\Delta$ is the scaling dimension of $\mathcal{V}$, and $|g|^{\frac{1}{2-\Delta}}$ is the mass scale of the system.\par
When the spatial dimension is compact with length $L$, $E_{cut}$ can be written explicitly as

\begin{equation}
\label{eq:TCSA Ecut}
E_{cut}={\frac{2\pi}{L}N_{cut}}
\end{equation}

where $N_{cut}$ is the maximal conformal level in the CFT basis. In practice, it is more useful to work with the dimensionless terms $L|g|^{\frac{1}{2-\Delta}}$ and $N_{cut}$.\par
For the case of SG and DSG, $H_{CFT}$ was taken to be that of a free scalar boson in $d=2$ on a cylinder of length $L$. The perturbation is $\mathcal{V}_i=\cos(\beta_i \phi)$ with $i=1$ for SG and $i=1,2$ for DSG. The maximal value of $N_{cut}$ used in the calculations was 42 (which gives a spectrum of $\sim85,000$ energy levels), and the dimensionless term $l=ML$ with $M$ being the IR mass scale which is the breather mass (\ref{eq:quantum-breather-mass}) with $n=1$, showed best convergence when $l\sim1$ for SG and $l\sim3$ for DSG.

\section{Sine - Gordon and double Sine-Gordon} \label{SG-and-DSG}

In this appendix introduce Sine-Gordon (SG) and Double Sine-Gordon (DSG) as the QFT models which will be tested with the measures from the previous sections.

\subsection{SG model}

Sine-Gordon (SG) is a known $(1+1)$D field theory which is well studied throught the years, describing physical phenomena such as Josephson junctions \cite{Josephson} and $1$D spin chains. 
It is a non-linear integrable model with known non-perturbative solutions and their interactions.
The lagrangian density is

\begin{equation}
\label{eq:SG-lagrangian}
\mathcal{L}_{SG}=\frac{1}{2}\partial_\mu \phi \partial^{\mu}\phi - \mu_0\cos{\beta \phi}
\end{equation}

The basic soliton solutions are called kinks

\begin{equation}
\label{eq:kink-solution}
\phi(x,t)=4\arctan\left[\exp{\left(\pm\frac{x-vt}{\sqrt{1-v^2}}\right)}\right]
\end{equation}

and are thought of as centered around $x$ for which $\phi=0$ with a velocity $0\leq v < 1$ and a soliton doublet 'mass' of $$m=\frac{8\sqrt{\mu_0}}{\beta}$$ They correspond to an overall shift in the phase $\phi\rightarrow\phi+2\pi$ for the solutions with $+$ in the exponent, and $\phi\rightarrow\phi-2\pi$ for those with a minus sign, called 'anti-kinks'. These solutions are topological and carry a topological charge of $\theta_k=+1$ for the kinks and $\theta_k=-1$ for the anti-kinks.
A bound state of a kink and and anti-kink is called a 'breather'

\begin{equation}
\label{eq:SG-Breather}
\phi(x,t)=4\arctan\left[\exp{\left(\frac{\sqrt{1-\omega^2}\cos\omega t)}{\omega\cosh{\left(\sqrt{1-\omega^2}x\right)}}\right)}\right]
\end{equation}

with a continuous mass in $|\omega|<1$, that is calculated by inserting the solution to the enregy-momentum tensor.

\begin{equation}
\label{eq:SG-breather-mass}
M=2m\sin{\frac{\pi}{2}\omega}
\end{equation}

Since we are interested in the discrete spectrum of the model, for the quantized version of SG and later DSG it will be useful to look at the free field hamiltonian as a CFT of a free boson with $c=1$ and add the potential terms as a perturbation to the CFT

\begin{equation}
\label{eq:QSG-Hamiltonian}
H_{SG}=H_{CFT}+\mu \int dx:\cos\beta\phi:
\end{equation}

where $$H_{CFT}=\int dx :\frac{1}{2}(\partial_t\phi)^2 + \frac{1}{2}(\partial_x\phi)^2:$$

The quantum spectrum consists of a soliton with mass $M$ that relates to the coupling parameter $\mu$ \cite{Baxter} as 

\begin{equation}
\label{eq:mu-coupling-QSG}
\mu=\frac{2\Gamma(\Delta)}{\pi\Gamma(1-\Delta)}\left(\frac{\sqrt{\pi}\Gamma\left(\frac{1}{2-2\Delta}\right) M}{2\Gamma\left(\frac{\Delta}{2-2\Delta}\right)}\right)^{2-2\Delta}
\end{equation}

with $\Delta=\beta^2/8\pi$.
The breather mass is now quantized, and reads

\begin{equation}
\label{eq:quantum-breather-mass}
M_n=2M\sin\left(\frac{\pi}{2}n\frac{\Delta}{1-\Delta}\right)
\end{equation}

where $n=1,2,\dots,\Delta^{-1}-1$. It was discovered by Coleman \cite{Coleman} that for $\Delta<1$ SG is dual to the charge zero sector massive Thirring model, with the special case of $\Delta=\frac{1}{2}$ corresponding to a free massive Dirac field.

\subsection{DSG}

The Hamiltonian of the quantum DSG can be written in the same way as that of SG (\ref{eq:QSG-Hamiltonian}) with another potential term

\begin{equation}
\label{eq:QDSG-Hamiltonian}
H_{DSG}=H_{CFT}+\mu \int dx:\cos\beta\phi:+\lambda \int dx:\cos\left(\alpha\phi+\delta\right):
\end{equation}

where the new term is seen as a perturbation of (\ref{eq:QSG-Hamiltonian}), from which it is possible to calculate the perturbative form factors (PFFT) and add them to the integrable spectrum, using the dimensionless parameter 

\begin{equation}
\label{eq:QDSG-perturbation-parameter}
\eta\equiv\lambda\mu^{-\frac{1-\Delta_\alpha}{1-\Delta_\beta}}=\lambda\mu^{\frac{8\pi-\alpha^2}{8\pi-\beta^2}}
\end{equation}

In the regimes where the limits of $\eta\rightarrow0$ and $\eta\rightarrow\infty$ converge. The latter obtained by swapping the roles of the two potential terms such that the term with $\alpha$ is now part of SG and the term with $\beta$ becomes the perturbation.\par
It is worth mentioning that adding the second frequency breaks the degeneracy of the vacuum state explicitly, and when the frequencies $\beta$ and $\alpha$ are incommensurate, the vacuum of the theory becomes unique. As for the solitons and anti-solitons, breaking the $2\pi/\beta$ periodicity and lifting the degeneracy of the minimal state leads to an attractive linear potential between the soliton-antisoliton pairs which causes their collapse into a string of bound states \cite{DelfinoMussardo}.\par
Explicit calculations of the PFFT corrections for the vacuum energy, first and second breather masses and the S matrix are found in \cite{Banjok}. A semi-classical calculation gives the following correction to the first breather mass

\begin{equation}
\label{eq:QDSG-SC-first-breather-correction}
{M_1}^{k}=\sqrt{\mu}\beta + \lambda\frac{\alpha^2}{2\sqrt{\mu}\beta}\cos{\left(\frac{2\pi}{\beta}k+\delta\right)} + O(\lambda^2)
\end{equation}

Where $k$ is one of the $n$ copies of the breather due to the remaining vacuum degeneracy when $\alpha/\beta$ is rational.
\section{Random Matrix Theory} \label{RMT}

The theory of random matrices, as the name suggests, deals with matrices which are composed of random variables. When these matrices are symmetric under rotations (i.e. orthogonal or unitary transformations), and when the random variables are made of identical distributions that are independent up to the rotational symmetry, those are called Wigner matrices, and the hold a special place in quantum chaos since they were derived as an attempt to capture the complex behavior of 
quantum many-body problems, specifically heavy nuclei.
It is conjectured in what is now called the Wigner-Dyson-Gaudin-Mehta conjecture, that the spectral statistics of Wigner matrices are universal, i.e. are independent of the choice of distribution for the random variables in the matrix \cite{ErdosYau}. \par
Let us consider an $N\times N$ matrix $H$ where each entry is a random variable with the same probability distribution, with a zero mean and a variance of $1/N$

\begin{align} \label{eq:Wigner-matirces-moments}
\mathbb{E}[H_{ij}]&=0      &  \mathbb{E}[{|H_{ij}|}^2]&=\frac{1}{N}    &  \text{for} i,j=1,2,\dots N
\end{align}

The object of interest will be the joint probability distribution of these variables $P(H)$, which according to Euler's lemma are weighted by $\exp{-Tr(H^{2n})}$ for integer $n$. When rotated to the diagonalized basis the joint probability distribution is

\begin{equation}
\label{eq:eigenbasis-JPD}
P(\lambda_1,\lambda_2,\dots,\lambda_N) = C_N \Delta^{\beta}\exp{\{-\beta N V(\lambda)\}}
\end{equation}

Where $C_N$ is a normalization constant, $\beta$ is the dyson index, and it indicates rotational symmetry of $H$:  $\beta=1$ for real symmetric matrices, $\beta=2$ for complex hermitian matrices and $\beta=4$ for quaternion self-adjoint matrices. $\Delta=\prod_{j<i}(\lambda_i-\lambda_j)$ is the Vandermonde determinant and it is common to choose a quadratic form for the potential and call the family of such matrices the Gaussian ensemble (these correspond to Gaussian random variables).\par
This joint probability distribution carries the whole of the information of the system. In order to understand the system, it is useful to look at the $n$ marginal probability distribution where $n<N$, also called the $n$-point correlation function

\begin{equation}
\begin{split}
R_n(\lambda_1,\lambda_2,\dots,\lambda_n)
&= \frac{N!}{(N-n)!}
\int P(\lambda_1,\lambda_2,\dots,\lambda_N) \\
&\qquad \times d\lambda_{n+1}\, d\lambda_{n+2}\cdots d\lambda_N .
\end{split}
\label{eq:n-point-correlation}
\end{equation}

Where the combinatorical factor outside of the integration counts the number of possible $N-n$ integration combinations, assuming the set of eigenvalues is ordered.\par
For example, the spectral level density, which is defined generally as

\begin{equation}
\label{eq:spectral-density}
\rho(\lambda)=\frac{1}{N} \sum_i \delta(\lambda-\lambda_i)
\end{equation}

Is the 1-point correlation $R_1(\lambda)$. There are several methods to derive the spectral density explicitly, most include for simplicity the large $N$ limit. Wigner originally used the moments method of computing $\mathbb{E}[Tr(H^n)]$. Another method is first writing the Gaussian ensemble version of (\ref{eq:eigenbasis-JPD}) in exponential form

\begin{equation}
\begin{split}
P(\lambda_1,\dots,\lambda_N)
&= C_N \exp\Biggl\{
-\frac{\beta N}{2}\sum_i \lambda_i^2
+ \beta \sum_{i<j}\log\bigl|\lambda_i-\lambda_j\bigr|
\Biggr\}.
\end{split}
\label{eq:eigenbasis-JPD-exp}
\end{equation}

Which now resembles the potential terms of a one dimensional constrained 2D coloumb gas in a harmonic potential. Taking the large $N$ limit is now similar to taking the semi-classical limit, where sums are replaced with integrals that include the spectral density $$\sum_i\rightarrow N\int d\lambda \rho(\lambda)$$ The requirement of a stationatry phase is now translated to the equation

\begin{equation}
\label{eq:Stationary-phase}
\lambda_i=\frac{1}{2}\int \frac{d\lambda\rho(\lambda)}{\lambda-\lambda_i}
\end{equation}

Which can be solved with the resolvent method. The solution is the renowned 'Wigner semi-circle law'

\begin{equation}
\label{eq:rho_0}
\rho_0(\lambda) = \frac{1}{\beta\pi} \sqrt{2\beta-\lambda^2}
\end{equation}

The 2-point and higher correlation function are delat with in \ref{pair-correlations}.\par
One can also derive the distribution of eigenvectors, based on the rotational invariance\cite{Reichl}. For a real symmetric $N\times N$ matrix, the $N$ dimensional eigenvectors are placed on an $N$ dimensional unit sphere, and cover the unit sphere uniformly as we move through the ensemble. The normalized joint probability distribution for its components is

\begin{equation}
\label{eq:joint-eigenvector-probability}
P(a_1,a_2,\dots,a_N) =\pi^{-N/2}\Gamma\left(\frac{N}{2}\right)\delta\left(\sum_i^N {a_i}^2-1\right)
\end{equation}

which we can use to obtain the probability density of a single component $y=a_i^2$

\begin{equation}
\begin{split}
P(y)
&= \int P(a_1,a_2,\dots,a_N)\,
\delta\!\left(y-a_i^2\right)\,
da_1\,da_2\cdots da_N \\
&= \frac{1}{\sqrt{\pi}}\,
\frac{\Gamma(N/2)}{\Gamma\!\bigl((N-2)/2\bigr)}\,
\frac{(1-y)^{(N-3)/2}}{\sqrt{y}} .
\end{split}
\label{eq:eigenvector-component}
\end{equation}

taking the large $N$ limit and changing variables to $\eta=Ny$, we get

\begin{equation}
\label{eq:Porter-Thomas}
P(\eta)=\frac{1}{\sqrt{2\pi\eta}}e^{-\frac{\eta}{2}}
\end{equation}

Which is known as the Porter-Thomas distribution. Similar results can be obtained for the GUE and GSE ensembles, assuming the components of the eigenvectors are complex or quaternion.\par
RMT can be generalized in various ways, for example beyond rotational invariance, where the joint probability distribution could be written with a general potential $V(\lambda)$, such as the Laguerre potential $V(\lambda)=a\log\lambda-\lambda/2$ of Wishart ensembles. The parameter $\beta$ can also be generalized to be continuous $\beta>0$. In this case it is shown in \cite{Dumitriu} that the corresponding random matrices are tri-diagonal, and for the Gaussian case (with a quadratic potential), this tri-diagonal matrices are

\begin{equation}
\label{eq:tri-diagonal}
H_\beta=\frac{1}{\sqrt{2}}
\begin{pmatrix}
N(0,2) & \chi_{(n-1)\beta} & & & \\
\chi_{(n-1)\beta} & N(0,2) & \chi_{(n-2)\beta} & & \\
 & \ddots & \ddots & \ddots & \\
 & & \chi_{2\beta} & N(0,2) & \chi_\beta\\
 & & & \chi_{\beta} & N(0,2)\\
\end{pmatrix}
\end{equation}

Where $N(0,2)$ are random variables of a normal distribution with $\mu=0$ and $\sigma=2$ and the PDF of $\frac{\chi_k}{\sqrt{2}}$ is $$\frac{2}{\Gamma(k/2)}y^{k-1}e^{-y^2}$$ .

\end{document}